\documentclass{siamltex}

\usepackage[breaklinks,colorlinks=true,linkcolor=blue,citecolor=red,backref=page]{hyperref}

\usepackage{amsfonts} 
\usepackage{amssymb,amsmath} 
\usepackage{graphicx,color}
\usepackage{fancybox}
\graphicspath{{./Fig/}}

\DeclareMathAlphabet{\itbf}{OML}{cmm}{b}{it}
 \DeclareMathAlphabet\mathbfcal{OMS}{cmsy}{b}{n}

\renewcommand{\hat}{\widehat}
\renewcommand{\tilde}{\widetilde}
\def\RR{\mathbb{R}}
\def\EE{\mathbb{E}}
\def\TT{\mathbb{T}}

\def\bx{{{\itbf x}}}

\def\bnu{\boldsymbol{\nu}}

\def\om{\omega}
\def\la{\lambda}

\def\bK{{\itbf{K}}}

\def\ba{{\itbf{a}}}
\def\bb{{\itbf{b}}}

\def\bI{{\bf I}}

\def\cP{\boldsymbol{\mathcal{P}}}

\def\cT{{\mathcal{T}}}

\def\cbP{\boldsymbol{\mathcal{P}}}
\def\cbT{\boldsymbol{\mathcal{T}}}
\def\cbR{\boldsymbol{\mathcal{R}}}

\def\bH{{\bf H}}
\def\bK{{\bf K}}

\newtheorem{lem}{Lemma}
\newtheorem{thm}{Theorem}
\newtheorem{prop}{Proposition}

\title{Enhanced wave transmission in random media with mirror symmetry} 

\author{
Liliana Borcea\footnotemark[1]
\and
Josselin Garnier\footnotemark[2]
}

\begin{document}

\maketitle

\renewcommand{\thefootnote}{\fnsymbol{footnote}}

\footnotetext[1]{Department of Mathematics, University of Michigan,
  Ann Arbor, MI 48109-1043 ({borcea@umich.edu})}  
\footnotetext[2]{Centre de Math\'{e}matiques Appliqu\'{e}es, Ecole Polytechnique, Institut Polytechnique de Paris, 91120 Palaiseau, France ({josselin.garnier@polytechnique.edu})}

\begin{abstract}
We present an analysis of enhanced wave transmission through random media  with mirror symmetry about a reflecting barrier. The mathematical model is  
the acoustic wave equation and we consider two setups, where the wave propagation is along a preferred direction:  in a randomly layered medium and  in a randomly perturbed waveguide. We use the asymptotic stochastic  theory of wave propagation in random media to characterize the statistical moments of the frequency-dependent random transmission and reflection 
coefficients, which are scalar-valued in layered media and matrix-valued in waveguides.  With these moments, we can quantify explicitly  the enhancement of the   net mean transmitted intensity, induced by wave interference near the barrier.
  
\end{abstract}

\section{Introduction}
\label{sect:intro}
Multiple scattering of waves traveling through disordered media is a serious impediment for applications like imaging and free space communications. 
This has motivated the pursuit of strategies for wave transmission enhancement and mitigation of scattering effects. 

At propagation distances (depths)  that do not exceed a few scattering mean free paths, the wave field retains some coherence. Mitigation 
strategies seek to enhance this coherence by:  filtering  the incoherent wave components, like  in optical coherence tomography \cite{huang1991optical} and in imaging in waveguides 
with rough boundary \cite{borcea2013quantitative};  correcting wavefront distortion in adaptive optics \cite{hardy1998adaptive};  or  using coherent interferometry 
\cite{borcea2006adaptive,borcea2021imaging}.

Beyond a few scattering mean free paths, the wave field is incoherent and it is typically described by the radiative transfer theory \cite{chandrasekhar1960radiative,ryzhik1996transport,borcea2016derivation} or the diffusion theory \cite{van1999multiple}.
These theories neglect wave interference effects that cause phenomena like coherent backscattering enhancement a.k.a. weak localization 
\cite{wolf1985weak,garnier2021enhanced} and Anderson  localization \cite{lagendijk2009fifty}. 
Such interference effects can be exploited 
for enhancing transmission through a strongly scattering medium.  In \cite{dorokhov1984coexistence} it was shown, using random matrix theory,  that in a disordered three-dimensional metallic body,  some of the eigenvalues of the transmission matrix are close to one. The eigenvectors for such  eigenvalues are known as open channels and their existence has been demonstrated in optics experiments in 
 \cite{vellekoop2008universal}. If the transmission matrix can be measured,
the open channels can be determined and used for improved focusing and delivering waves deep inside disordered media
\cite{vellekoop2008universal,popoff2014coherent,cao2022shaping}.

Recent developments show that interesting wave interference phenomena can also be induced by mirror symmetry in  chaotic cavities and  in waveguides filled with disordered media. Large conductance enhancement through a reflecting barrier has been demonstrated in  \cite{whitney2009huge} for symmetric quantum dots and  in \cite{gopar1996invariant,gopar2006transport} for symmetric chaotic cavities. Experimental demonstration of broadband wave transmission enhancement through diffusive, symmetric slabs  with a barrier in the middle is given  in \cite{cheron2019broadband,davy2021experimental}.
Symmetric media are also encountered when studying 
waves propagating in a random half-space with Dirichlet boundary condition. The method of images  replaces this half-space problem  by a full-space problem with symmetric sources and media \cite{messaoudi2023boundary}.

Our goal in this paper is to study mathematically wave transmission enhancement in disordered systems with mirror symmetry. The analysis can be carried out for any type of linear waves, but for simplicity we consider acoustic waves. We are interested in two setups,  where the wave propagation is along a preferred direction: a randomly layered medium and a waveguide filled with a disordered medium. In both cases, the disordered medium is modeled by random fluctuations of the coefficients of the wave equation. These fluctuations are mirror symmetric with respect to a reflecting barrier. The interaction of the waves with the barrier and 
the random medium is described by frequency-dependent reflection and transmission coefficients, which are scalar-valued in 
the layered case and matrix-valued in waveguides. We use the stochastic asymptotic theory of wave propagation 
\cite{fouque2007wave, garnier2008effective} to write  the statistical moments of these coefficients and thus quantify explicitly the 
net mean transmission intensity for various opacities of the barrier.  In both settings we find that the mirror symmetry has a beneficial effect on the transmission. This effect is more striking when the obstruction at the barrier is strong.

We organize the analysis and results in two main sections: We begin in section \ref{sect:1D} with the case of waves propagating at normal incidence through a randomly layered medium. This lets us  introduce the main ideas  in a simpler, one-dimensional setting,  so we can analyze the enhanced transmission in great detail. Then, we study in section 
\ref{sect:Waveg} transmission through random waveguides. The statistical moments of the reflection and transmission matrices 
in random waveguides are known, but their computation is much more complicated  than in the one-dimensional case \cite{garnier2008effective}.  Thus, for waveguides, we consider a regime of weak scattering in the random medium, so we can get an explicit approximation of the net transmitted intensity.  The involved calculations needed to derive the results in sections \ref{sect:1D} and \ref{sect:Waveg} are given in appendixes. We end with a summary in section \ref{sect:sum}.

\section{Enhanced transmission in randomly layered media}
\label{sect:1D}
We give here the analysis of wave transmission in one-dimensional (layered) random media 
with mirror symmetry. We begin in section \ref{sect:1D_1} with the setup and the wave decomposition in 
forward and backward going modes. The analysis of the reflection and transmission of these modes at the reflecting barrier is in section \ref{sect:1D_2} and  in the random medium is in section \ref{sect:1D_3}. We gather the results in section 
\ref{sect:1D_4} to quantify the transmission enhancement.

\subsection{Setup}
\label{sect:1D_1} 
One-dimensional wave propagation along the $z-$axis is described by the first order system 
\begin{align}
\left[\begin{pmatrix} 
\rho(z) & 0 \\
0 & K^{-1}(z)
\end{pmatrix}
\partial_t + \begin{pmatrix} 
0 & 1 \\
1 & 0 
\end{pmatrix}  \partial_z \right]
 \begin{pmatrix} u(t,z) \\
p(t,z) \end{pmatrix} = {\bf 0}, \qquad t \in \RR, ~~ z \in \RR,
\label{eq:1D.1}
\end{align}
where  $p$ is the acoustic pressure and $u$ is the velocity. The medium is modeled by 
the variable density $\rho$ and bulk modulus $K$, which determine the local wave speed $c$ and impedance 
$\zeta$, 
\begin{equation}
c(z) = \sqrt{K(z)/\rho(z)}, \quad 
\zeta(z) = \sqrt{K(z) \rho(z)}.
\label{eq:def_czeta}
\end{equation}

The medium contains a thin barrier at $z \in (-d/2,d/2)$, sandwiched  between
two randomly perturbed, symmetric regions at $d/2 \le |z| \le L$. Assuming that the $z-$axis is horizontal,
we call the region $z < -d/2$ the left side of the barrier and the region $z > d/2$ the right side of the barrier. 
The medium is modeled by 
\begin{align}
\hspace{-0.13in}\rho(z)  = 
\left\{
\begin{array}{ll}
\rho_0 &\mbox{ if } |z| \ge {d}/{2},\\ 
\rho_1 &\mbox{ if } |z|< {d}/{2} ,
\end{array}\right.  \mbox{and} ~
\frac{1}{K(z)}  = 
\left\{
\begin{array}{ll}
\frac{1}{K_0} & \mbox{ if } |z|>L,\\
\frac{1}{K_0}  \big[1+\mu(|z|) \big]  & \mbox{ if }  |z|\in \big[d/2,L\big] ,\\
\frac{1}{K_1} &\mbox{ if } |z|<{d}/{2} ,
\end{array} \right.
\label{eq:1D.3}
\end{align}
where $\rho_j$ and $K_j$ are positive constants, for $j = 0,1$ and $\mu$  is a mean zero, mixing  random process, 
satisfying the uniform bound  $|\mu| < 1$, so that the bulk modulus is a positive function \cite[Chapter 6]{fouque2007wave}.
Note that only the bulk modulus has random fluctuations in our model. This simplifies the presentation and unifies it with that in the next section, because \eqref{eq:1D.1} 
reduces to the standard second-order wave equation for the pressure at $|z| > d/2$ and at $|z| < d/2$.  Random fluctuations of the density can be included, and the results are qualitatively the same \cite[Chapter 17]{fouque2007wave}.

The interaction of the waves with the medium depends on frequency, so 
we Fourier transform  with respect to time,
\begin{equation}
\hat p(\om,z) = \int_{\RR} d t \, e^{i \om t} p(t,z), \quad 
\hat u(\om,z) = \int_{\RR} d t \, e^{i \om t} u(t,z),
\label{eq:1D.FT}
\end{equation}
and then decompose the wave field  into right (forward) going and left (backward) going modes \cite[Chapter 7]{fouque2007wave}. The decomposition at $|z|  \ge d/2$ is 
\begin{align}
\hat a(\om,z) &= \left[\zeta_0^{-1/2} \hat p(\om,z) +\zeta_0^{1/2} \hat u(\om,z)\right] e^{-i \om \frac{z}{c_o} },\label{eq:1D.4}\\
\hat b(\om,z) &= \left[-\zeta_0^{-1/2} \hat p(\om,z) +\zeta_0^{1/2} \hat u(\om,z)\right] e^{i \om \frac{z}{c_o} },
\label{eq:1D.5}
\end{align}
where $c_0 = \sqrt{K_0/\rho_0}$ and $\zeta_0 = \sqrt{K_0 \rho_0}$. The decomposition at $|z| < d/2$ 
is similar, except that  $c_0$ and $\zeta_0$ are replaced by $c_1 = \sqrt{K_1/\rho_1}$ and $\zeta_1 = \sqrt{K_1 \rho_1}$. Note that equations (\ref{eq:1D.FT}-\ref{eq:1D.5}) give 
\begin{align}
p(t,z) &=\frac{\zeta_0^{{1}/{2}}}{4 \pi}  \int_{\RR} d \om \, e^{-i \om t} 
\left[ \hat a(\om,z) e^{i \om \frac{z}{c_o}} - \hat b(\om,z)  e^{-i \om \frac{z}{c_o}}\right], \label{eq:modedec1}\\
u(t,z) &=\frac{\zeta_0^{-{1}/{2}}}{4 \pi}  \int_{\RR} d \om \, e^{-i \om t} 
\left[ \hat a(\om,z) e^{i \om \frac{z}{c_o}} + \hat b(\om,z)  e^{-i \om \frac{z}{c_o}}\right] .\label{eq:modedec2}\
\end{align}
This is a decomposition in monochromatic waves propagating along the $z-$axis in the right  direction, 
with amplitude $\hat a$, and the left  direction, with amplitude $\hat b$. 

The wave excitation specifies $
\hat a(\om,-L),$ and corresponds to a wave impinging on the heterogeneous medium at $z = -L$. The goal is to quantify the wave emerging  at $z = L$, with amplitude $\hat a(\om,L)$ (see Fig.~\ref{fig:3sect}).  Since the medium is homogeneous at $z > L$, 
the wave is outgoing there i.e.,  $\hat b(\om,z) = \hat b(\om,L) = 0$ for $z \ge L$.

\subsection{Model of the barrier}
\label{sect:1D_2} 
The mapping of the wave modes on the left of  the barrier, at $z = -d/2$,  to the modes on the right of the barrier, at  $z = d/2$, is given by the $2 \times 2$ frequency-dependent propagator matrix ${\bf P}_1$. The expression of this matrix is 
derived in   Appendix \ref{ap:1DAp.1}, by imposing the continuity of 
the pressure and velocity at $z = \pm d/2$. We state the result in the next lemma.

\vspace{0.05in}\begin{lemma}
\label{lem.1}
We have
\begin{equation}
\begin{pmatrix} 
\hat{a}(\omega,d/2)  \\
\hat{b} (\omega,d/2)  
\end{pmatrix} = {\bf P}_1(\omega)
\begin{pmatrix} 
\hat{a}(\omega,-d/2)  \\
\hat{b} (\omega,-d/2) 
\end{pmatrix} ,\qquad 
{\bf P}_1(\omega) = 
 \begin{pmatrix} 
\alpha(\omega) & \overline{\gamma(\omega)}  \\
 \gamma(\omega)   &  \overline{\alpha(\omega)} 
\end{pmatrix}  ,
\label{eq:defPb}
\end{equation}
where the bar denotes complex conjugate and
\begin{align}
\alpha(\omega) &= \left[ \cos \Big(\frac{\omega d}{c_1} \Big) + \frac{i}{2} \Big(\frac{\zeta_1}{\zeta_0} + \frac{\zeta_0}{\zeta_1} 
\Big) \sin \Big(\frac{\omega d}{c_1} \Big)\right]   e^{-i \omega d/c_0}, \label{eq:defalphab}\\
\gamma(\omega) &= \frac{i}{2} \Big(\frac{\zeta_0}{\zeta_1} - \frac{\zeta_1}{\zeta_0}
\Big)\sin \Big(\frac{\omega d}{c_1} \Big).
\label{eq:defbetab}
\end{align}
\end{lemma}

The scattering matrix $ {\bf S}_1$ maps the wave mode amplitudes that impinge on the barrier to the outgoing wave mode amplitudes
\begin{equation}
\begin{pmatrix} 
\hat{a}(\omega,d/2)  \\
\hat{b} (\omega,-d/2)  
\end{pmatrix} = {\bf S}_1(\omega)
\begin{pmatrix} 
\hat{a}(\omega,-d/2)  \\
\hat{b} (\omega,d/2) 
\end{pmatrix}.
\end{equation}
Its expression follows from equation \eqref{eq:defPb},
\begin{equation} 
{\bf S}_1(\omega) = 
 \begin{pmatrix} 
T_1(\omega) &  {R}_1(\omega) \\
R_1(\omega)   &  T_1(\omega), 
\end{pmatrix}, \quad R_1(\omega) = -\frac{\gamma(\omega)}{\overline{\alpha(\omega)}},\quad
T_1(\omega) = \frac{1}{\overline{\alpha(\omega)}},
\label{eq:transmB}
\end{equation}
where $T_1$ and $R_1$ are the transmission and reflection coefficients of the barrier.

We are interested in  a thin barrier, with $d$ much smaller than the wavelength, so there are no trapped propagating modes at $z \in (-d/2,d/2)$. There are two distinguished 
asymptotic regimes that give an order one net effect of the barrier: 
\\
\textbf{1:} The first regime is 
\begin{equation}
\frac{\om d}{c_j} \to 0 ~~ \mbox{for} ~ j = 0,1 ~~ \mbox{and} ~\frac{\zeta_0}{\zeta_1} \to \infty \quad \mbox{such that} \quad 
\frac{\zeta_0}{\zeta_1} \frac{\om d}{2 c_1} \to q(\om),
\label{eq:Reg1}
\end{equation}
with finite $q$. The asymptotic limit of the transmission and reflection coefficients is 
\begin{equation}
T_1(\om) = \frac{i}{i + q(\om)} \quad \mbox{and} 
\quad R_1(\om) = \frac{q(\om)}{i + q(\om)}.
\label{eq:Reg2}
\end{equation}
\\
\textbf{2:} The second regime is 
\begin{equation}
\frac{\om d}{c_j} \to 0 ~~ \mbox{for} ~ j = 0,1 ~~ \mbox{and} ~\frac{\zeta_1}{\zeta_0} \to \infty \quad \mbox{such that} \quad 
\frac{\zeta_1}{\zeta_0} \frac{\om d}{2 c_1} \to q(\om),
\label{eq:Reg3}
\end{equation}
and the transmission and reflection coefficients are 
\begin{equation}
T_1(\om) = \frac{i}{i + q(\om)} \quad \mbox{and} 
\quad R_1(\om) = -\frac{q(\om)}{i + q(\om)}.
\label{eq:Reg4}
\end{equation}
The two cases are similar, so we consider henceforth the asymptotic regime 
\eqref{eq:Reg1}.

\subsection{Reflection and transmission in the random medium}
\label{sect:1D_3} 
The propagation of waves in randomly layered media is studied in detail in \cite{fouque2007wave}. We gather the 
relevant results from there and characterize the transmission through the random medium with mirror symmetry in 
the next lemma, proved in Appendix \ref{ap:1DAp.2}.

\vspace{0.05in}\begin{lemma}
\label{lem.2}
We have
\begin{equation}
\label{eq:lemrand:1}
\begin{pmatrix} 
\hat{a}(\omega,-d/2)  \\
\hat{b} (\omega,-L)  
\end{pmatrix} = {\bf S}_-(\omega)
\begin{pmatrix} 
\hat{a}(\omega,-L)  \\
\hat{b} (\omega,-d/2) 
\end{pmatrix} ,\qquad 
\begin{pmatrix} 
\hat{a}(\omega,L)  \\
\hat{b} (\omega,d/2)  
\end{pmatrix} = {\bf S}_+(\omega)
\begin{pmatrix} 
\hat{a}(\omega,d/2)  \\
\hat{b} (\omega,L) 
\end{pmatrix} ,
\end{equation}
where ${\bf S}_-$ and ${\bf S}_+$ are 
the scattering matrices of the first and second random regions 
$[-L,-d/2]$ and $[d/2,L]$, respectively:
\begin{equation}
{\bf S}_-(\omega) = 
 \begin{pmatrix} 
T_-(\omega) &  \tilde{R}_-(\omega) \\
R_-(\omega)   &  T_-(\omega)
\end{pmatrix} ,
\qquad
{\bf S}_+(\omega) = 
 \begin{pmatrix} 
T_+(\omega) &  \tilde{R}_+(\omega) \\
R_+(\omega)   &  T_+(\omega)
\end{pmatrix}.
\label{eq:lemrand:2}
\end{equation}
{Due to the symmetry of the random medium, the transmission and reflection coefficients in these matrices satisfy}
\begin{equation}
T_-(\om)=T_+(\om), \qquad R_-(\om)=\tilde{R}_+(\om), \qquad \tilde{R}_-(\om) = R_+(\om) .
\label{eq:lemrand:3}
\end{equation}
\end{lemma}

{For quantifying the net mean intensity transmitted through the medium we only need the expressions of the statistical moments of the square modulus of the transmission coefficient of one random section.
The moments of  $T_+$ are studied in \cite[Section 7.1.5]{fouque2007wave}
and the moments of  $R_+$ and $\tilde R_+$ are in \cite[Section 9.2.1]{fouque2007wave}, in the so-called strongly heterogeneous white-noise regime defined by the scaling relations
\begin{equation}
\ell_c \ll \lambda \ll L, \qquad {\rm Var}(\mu) = O(1),
\label{eq:regime1}
\end{equation}
where $\ell_c$ is the correlation length of the random fluctuations $\mu$ of the medium and  $\lambda=2\pi c_o/\om$ 
is the wavelength. If in addition we have ${\rm Var}(\mu)\ell_c L /\lambda^2 = O(1)$, then  the effect of the random medium  on the transmittivity is of order one. The expressions of the moments of $T_+$ are
\begin{equation}
\mathbb{E} \big[ |T_+(\om)|^{2n} \big] = \exp\Big(-\frac{ L}{4 L_{\rm loc}(\om)}\Big)
\int_0^\infty   e^{-  L s^2 / L_{\rm loc}(\om)} \frac{2 \pi s \sinh(\pi s)}{\cosh^2(\pi s)} \phi_n(s) ds,
\label{eq:lemrand:4}
\end{equation}
for any positive integer $n$.} Here  $\EE$ is the expectation with respect to the law of the process $\mu$, the functions $\phi_n$ are 
defined by 
\begin{equation}
\phi_1(s)=1,\qquad \phi_n(s)= \prod_{j=1}^{n-1} \frac{s^2 +(j-\frac{1}{2})^2}{j^2}, \quad n \geq 2,
\label{eq:defphin}
\end{equation}
and $L_{\rm loc}$ is the localization length of the random medium, which depends on the frequency $\om$ and the statistics of $\mu$:
\begin{equation}
\label{eq:defLloc}
\frac{1}{L_{\rm loc}(\om)} = \frac{\omega^2}{4 c_0^2} \int_\RR \EE[\mu(0)\mu(z)] dz  .
\end{equation}
Note that $1/L_{\rm loc}$ is of the order of ${\rm Var}(\mu)\ell_c /\lambda^2 $.
If the wave travels deep in the medium i.e., if $L \gg L_{\rm loc}$, then the moment formula \eqref{eq:lemrand:4} simplifies to 
\begin{equation}
\mathbb{E} \big[ |T_+(\om)|^{2n} \big] \simeq 
\frac{\pi^{5/2} }{2 [L/L_{\rm loc}(\om)]^{3/2}} \phi_n(0) \exp\Big[-\frac{ L}{4 L_{\rm loc}(\om)}\Big].
\label{eq:strongLayered}
\end{equation}

It is shown in \cite[Section 7.3]{fouque2007wave} that the moment formulas  (\ref{eq:lemrand:4}) also hold in the asymptotic regime where the correlation length of the medium is similar to the wavelength and smaller than the propagation distance, and the medium fluctuations have small relative amplitude,
\begin{equation}
\ell_c \sim \lambda \ll L,\qquad {\rm Var}(\mu)\ll 1.
\end{equation}
{In this so-called weakly heterogeneous regime,  the effect of the random medium fluctuations on the transmittivity is of order one when ${\rm Var}(\mu) \ell_c L/\la^2 = O(1)$.} The moments of $T_+$ are given by (\ref{eq:lemrand:4}) with the localization length 
$$
\frac{1}{L_{\rm loc}(\om)} = \frac{\omega^2}{4 c_0^2} \int_\RR \EE[\mu(0)\mu(z)] \cos\Big(\frac{2\omega z}{c_0}\Big) dz .
$$
This is slightly different from the expression (\ref{eq:defLloc})  of $L_{\rm loc}$ in the strongly heterogeneous white-noise regime (\ref{eq:regime1}).

\subsection{Transmission enhancement}
\label{sect:1D_4}

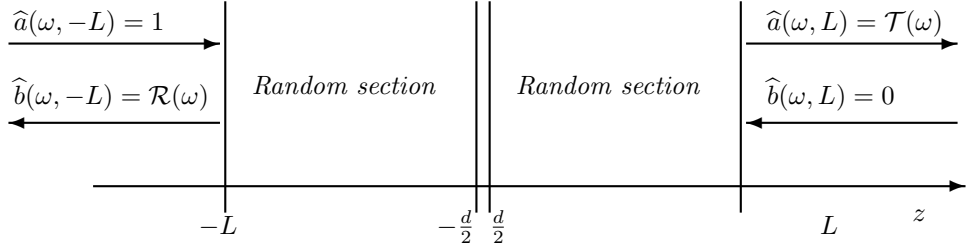
\begin{figure} 
\centerline{ 
\begin{picture}(300,75) 
\unitlength=1pt 
\thicklines 
\put(0,10){\vector(1,0){330}} 
\put(50,0){\line(0,1){80}} 
\put(245,0){\line(0,1){80}} 
\put(150,0){\line(0,1){80}} 
\put(145,0){\line(0,1){80}}
\put(40,-8){$-L$} 
\put(275,-8){$L$} 
\put(310,-3){$z$} 
\put(130,-8){$-\frac{d}{2}$} 
\put(150,-8){$\frac{d}{2}$} 
\put(60,45){{\it Random section}}
\put(160,45){{\it Random section}}
\put(48,34){\vector(-1,0){80}} 
\put(-30,41){$\hat b(\om,-L)={\cal R}(\om)$} 
\put(-32,64){\vector(1,0){80}} 
\put(-30,69){$\hat a(\om,-L)=1$} 
\put(327,34){\vector(-1,0){80}} 
\put(255,69){$\hat a(\om,L)={\cal T}(\om)$} 
\put(247,64){\vector(1,0){80}} 
\put(255,41){$\hat b(\om,L)=0$} 
\end{picture} 
} 
\vspace{0.1in}
\caption{Schematic of transmission and reflection in the random medium 
with mirror symmetry about the thin barrier located at $z \in (-d/2,d/2)$.}
\label{fig:3sect}
\end{figure}
We are interested in the transmission of the wave field through the medium, illustrated schematically in Fig. \ref{fig:3sect}. 
The result is stated in the next theorem, proved in Appendix \ref{ap:1DAp.3}. In light of Lemma \ref{lem.2}, 
we simplify the notation in its statement using
\begin{equation}
T(\om) = T_+(\om) = T_{-}(\om), \quad R(\om) = R_+(\om) = \tilde{R}_-(\om).
\label{eq:Tra2}
\end{equation}

\vspace{0.05in}
\begin{thm}
\label{prop.1}
The transmission coefficient of the system is 
\begin{equation}
\mathcal{T}(\om) = T^2(\om) T_1(\om) [ 1 - R(\om)]^{-1}\big[1 - \big(2R_1(\om)-1 \big) R(\om) \big]^{-1},
\label{eq:Tra1}
\end{equation}
and the expression of the mean transmitted intensity is 
\begin{equation}
\EE\big[ \big|\mathcal{T}(\om)\big|^2 \big] = \sum_{k=0}^\infty \tau_k(\omega)  \EE \left[ |T(\om)|^4\big(1 - |T(\om)|^2 \big)^k \right], \label{eq:meanint}
\end{equation}
where the moments of $T$ are given in equation \eqref{eq:lemrand:4} and 
\begin{equation}
\tau_k(\om) = \frac{1}{4}\left| 1 - \left(2 R_1(\om) -1 \right)^{k+1} \right|^2.
\label{eq:Tra5}
\end{equation}
\end{thm} 

\vspace{0.05in}Note that the coefficients \eqref{eq:Tra5} satisfy $\tau_k  \le 1$, because according to equation \eqref{eq:Reg2},  
\begin{equation}
\label{eq:Trab}
|2R_1(\om)-1| =\left|\frac{q(\om)-i}{q(\om) + i}\right| = 1.
\end{equation}
Using this inequality in  equation \eqref{eq:meanint}, we deduce  that\footnote{We can exchange the expectation with the sum because the series is uniformly convergent. } 
\begin{align}
\EE\big[ \big|\mathcal{T}(\om)\big|^2 \big] \le \EE \left[  |T(\om)|^4 \sum_{k=0}^\infty 
\big(1 - |T(\om)|^2 \big)^k \right] = \EE \left[  |T(\om)|^2\right].
\label{eq:mean1}
\end{align} 
Thus, no matter how weak or strong the barrier is, $\EE[|\mathcal{T}|^2]$ cannot exceed the mean intensity transmitted over half the distance, through one region of the random medium.

We analyze next the transmitted intensity in various scenarios. There are two extreme cases:

$\bullet $
The first extreme case is not interesting because it assumes no random fluctuations. We have $T = 1$  and the 
 transmitted intensity is deterministic and equal to the squared modulus of the transmission coefficient of the barrier
\begin{equation}
\big|\mathcal{T}(\om)\big|^2 \stackrel{\eqref{eq:meanint}}{=}\tau_0(\om) \stackrel{\eqref{eq:Tra5}}{=} \big| 1 - R_1(\om) \big|^2 \stackrel{\eqref{eq:Reg2}}{=}
\big|T_1(\om)\big|^2.\label{eq:mean2}
\end{equation}

\begin{figure}
\centering
\includegraphics[width=2.in]{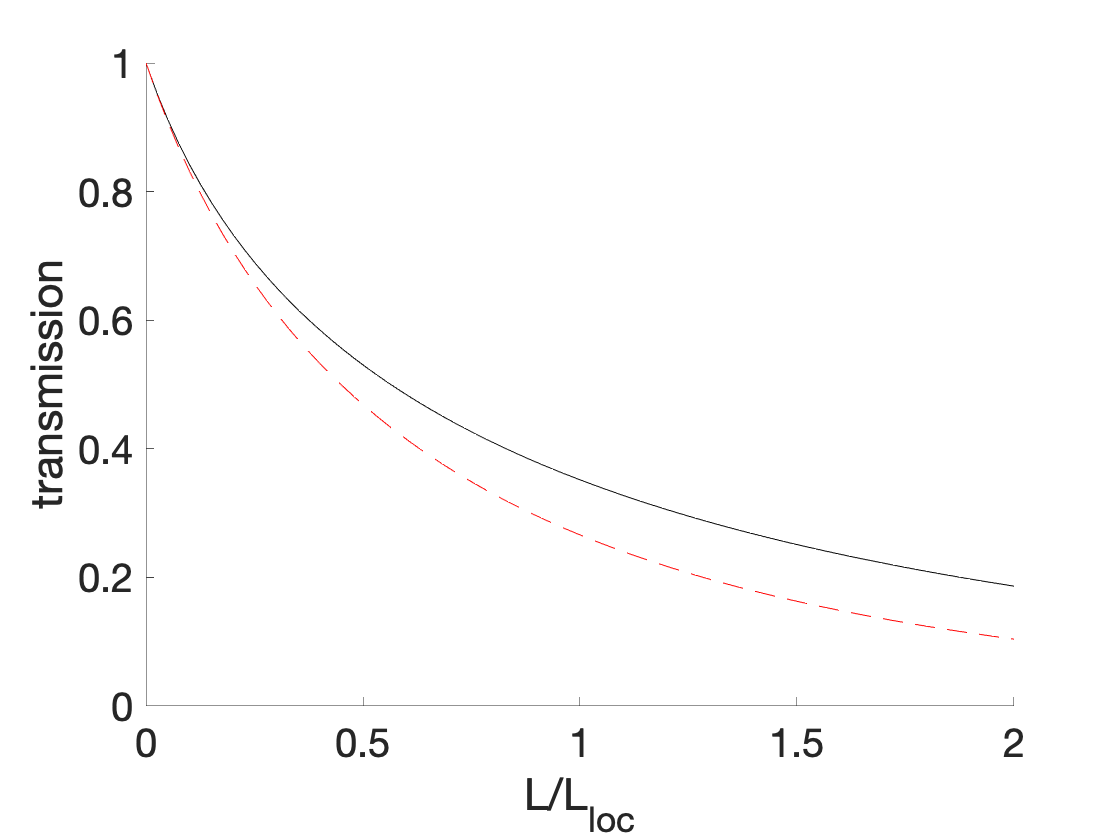} 
\includegraphics[width=2.in]{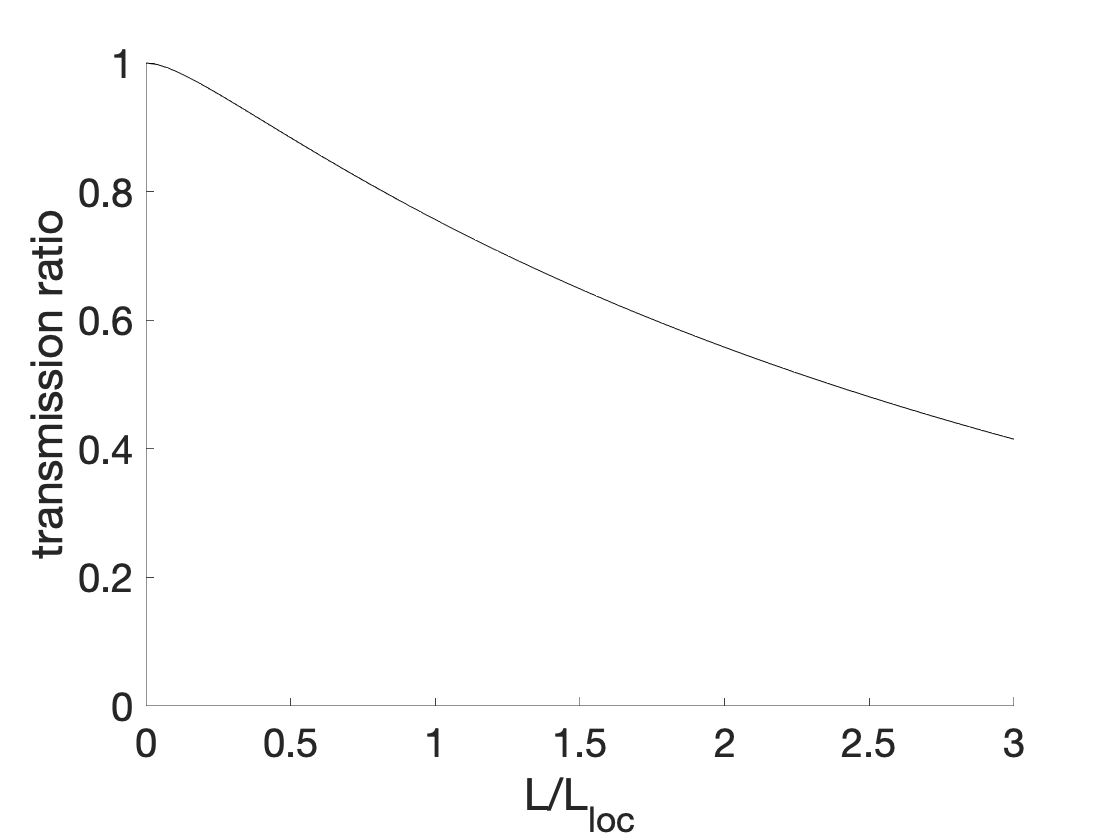} 
\caption{Mean transmitted intensity $\EE[ |\mathcal{T}|^2 ]$ of the system as a function of the strength $L/L_{\rm loc}$  of the randomly scattering medium in the absence of the barrier.  Left: The black solid line is the result \eqref{eq:mean3} for symmetric media and the red dashed line is the result \eqref{eq:mean5} for independent media. Right: Ratio of the mean transmission of independent media and the mean transmission of symmetric media.}
\label{fig:comp1}
\end{figure}

$\bullet $ The second extreme case assumes no barrier i.e., $T_1 = 1$ and $R_1 = 0$. Then, the coefficients \eqref{eq:Tra5} 
satisfy $\tau_k = 0$ for odd $k$ and $\tau_k = 1$ for even $k$, and the mean transmitted intensity is, from \eqref{eq:meanint},
\begin{align}  
\EE\big[ \big|\mathcal{T}(\om)\big|^2 \big] &=\EE \left[ |T(\om)|^4\sum_{k=0}^\infty \big(1 - |T(\om)|^2 \big)^{2k} \right]
=\EE \left[\frac{|T(\om)|^2}{2 - |T(\om)|^2}\right].
 \label{eq:mean3}
\end{align}
If in addition the random sections are strongly scattering, i.e. $L$ is larger than $L_{\rm loc}$
so the approximation \eqref{eq:strongLayered} holds, 
then we have 
\begin{equation}
\EE\big[ \big|\mathcal{T}(\om)\big|^2 \big]  \stackrel{\eqref{eq:mean3}}{=} \sum_{k=1}^\infty 2^{-k} \EE \left[ |T(\om)|^{2k}\right]
\stackrel{\eqref{eq:strongLayered}}{\simeq} C \EE \left[ |T(\om)|^{2}\right], 
\label{eq:mean4}
\end{equation}
where 
\begin{equation}
C = \sum_{k=1}^\infty 2^{-k} \phi_k(0) \approx 0.59.
\end{equation}
The result (\ref{eq:mean3}) says that, as expected from the estimate \eqref{eq:mean1}, the mean transmitted intensity 
through the symmetric random medium occupying the interval $[-L,L]$ is less than the intensity transmitted through 
the single region $z \in [0,L]$. However, the symmetry helps, because if the two random regions were independent, the mean transmitted intensity would be (see Appendix \ref{ap:1DAp.4}) 
\begin{equation}
\EE\big[ \big|\mathcal{T}(\om)\big|^2 \big] = \sum_{k=0}^\infty \left\{ \EE \left[ |T(\om)|^2 \big(1 - |T(\om)|^2 \big)^k \right] \right\}^2.
\label{eq:mean5}
\end{equation} 
This is smaller than \eqref{eq:mean3}, as illustrated in Fig. \ref{fig:comp1}. 

Physically, we 
can interpret the enhanced transmission due to symmetry as follows: It is known that the distribution of the random transmittivity has a small component close to one,  that actually gives the value of the mean transmittivity \cite[Section 7.1.6]{fouque2007wave}. The medium configurations that give transmittivity close to one are called open channels in the physics literature \cite{dorokhov1984coexistence,beenakker}.  Efficient transmission through two independent media of length $L$ requires the lucky situation where both media are open channels. For the symmetric case, this requires only one medium of length $L$ to be an open channel as the symmetric medium is then automatically an open channel.

Now we demonstrate the transmission enhancement in the presence of the barrier: 

\vspace{0.05in}
$\bullet$ First, we can see from equation \eqref{eq:meanint} that if the random sections are weakly scattering, i.e. $L/L_{\rm loc} \ll 1$,
then $\EE[|R|^2] = 1 -\EE[ |T|^2 ] \ll 1$ and
we can approximate the mean transmitted intensity by 
\begin{align*}
\EE[ |\mathcal{T}(\om)|^2 ] &= \tau_0(\om) \EE[|T(\om)|^4] + \tau_1(\om) \EE\left[[|T(\om)|^4 |R(\om)|^2 \right]
+ o \left(\EE[|R(\om)|^2]\right) \\
&= \tau_0(\om) + (\tau_1(\om)-2 \tau_0(\om)) \EE[|R(\om)|^2] + o \left(\EE[|R(\om)|^2]\right).
\end{align*}
Equation \eqref{eq:Tra5} gives 
\begin{align*}
\tau_0(\om) &= \left| 1 - R_1(\om)\right|^2 \stackrel{\eqref{eq:Reg2}}{=}  |T_1(\om)|^2 \\
\tau_1(\om) &= 4 |R_1(\om)|^2 \left| 1 - R_1(\om)\right|^2 = 4 |T_1(\om)|^2 (1 - |T_1(\om)|^2),
\end{align*}
so to leading order in the reflection coefficient we have 
\begin{equation}
\EE[ |\mathcal{T}(\om)|^2 ] \approx   |T_1(\om)|^2 \left\{ 1 + 2\left( 1- 2 |T_1(\om)|^2\right) \EE[|R(\om)|^2] \right\}.
\label{eq:mean6}
\end{equation}
This is larger than the transmission intensity of the barrier $  |T_1(\om)|^2$, as long as the barrier is reflecting enough i.e., 
for $|T_1(\om)| < 1/\sqrt{2}$.  

\begin{figure}
\begin{center}
\begin{tabular}{cc}
\includegraphics[width=2.in]{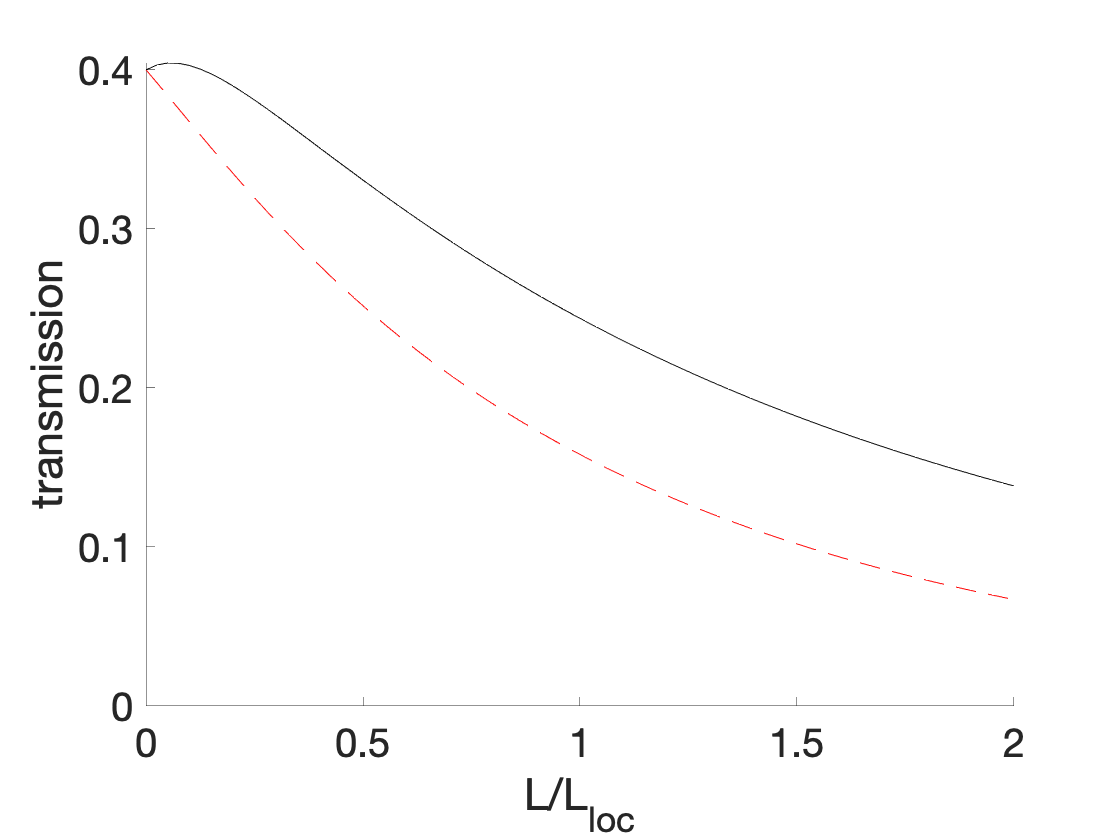} &
\includegraphics[width=2.in]{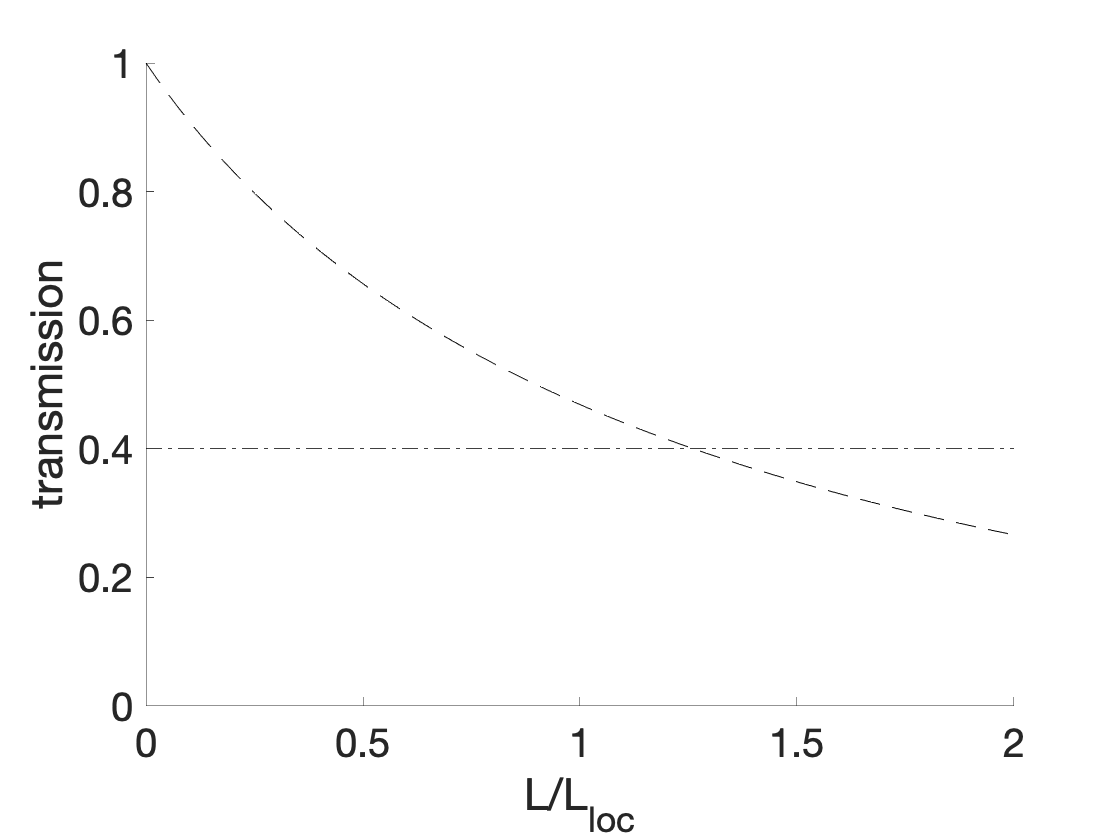}
\end{tabular}
\end{center}
\caption{Left: Mean transmission $\EE[|{\cal T}|^2]$ of the system as a function of the strength $L/L_{\rm loc}$ of the randomly scattering medium; the black solid line corresponds to the symmetric media and the red dashed line corresponds to the independent media. Right: the mean transmission $\EE[|T|^2]$ of one random section (dashed) and the transmission $|T_1|^2$ of the barrier (dot-dashed). Here $|T_1|^2=0.4$.}
\label{fig:comp2}
\end{figure}

$\bullet$ If the random sections are more scattering, i.e. $L\gtrsim L_{\rm loc}$ , then we must consider the series in \eqref{eq:meanint}. We compare 
the result in Fig. \ref{fig:comp2} with the mean intensity calculated in the absence of symmetry i.e., for two independent 
random media to the left and right of the barrier. The expression of the latter is
\begin{align}
\EE \big[ |{\cal T}(\om)|^2\big ] =  |T_1(\om)|^2 \sum_{k,k'=0}^\infty C_{k,k'}(\om)  \EE\big[|T(\om)|^2 (1-|T(\om)|^2)^k\big] \nonumber \\ 
\times \EE\big[|T(\om)|^2 (1-|T(\om)|^2)^{k'}\big],
\label{eq:meanIndep}
\end{align}
with 
$$
C_{k,k'}(\om)  = \left| \sum_{j=\max(k,k')}^{k+k'}
\frac{j!R_1^{2j-k-k'}(\om)  [1-2R_1(\om)]^{k+k'-j} }{(k+k'-j)! (j-k)!(j-k')!} 
\right|^2.
$$
Its derivation is given in Appendix \ref{ap:1DAp.4}.

\begin{figure}
\begin{center}
\begin{tabular}{cc}
\includegraphics[width=2.in]{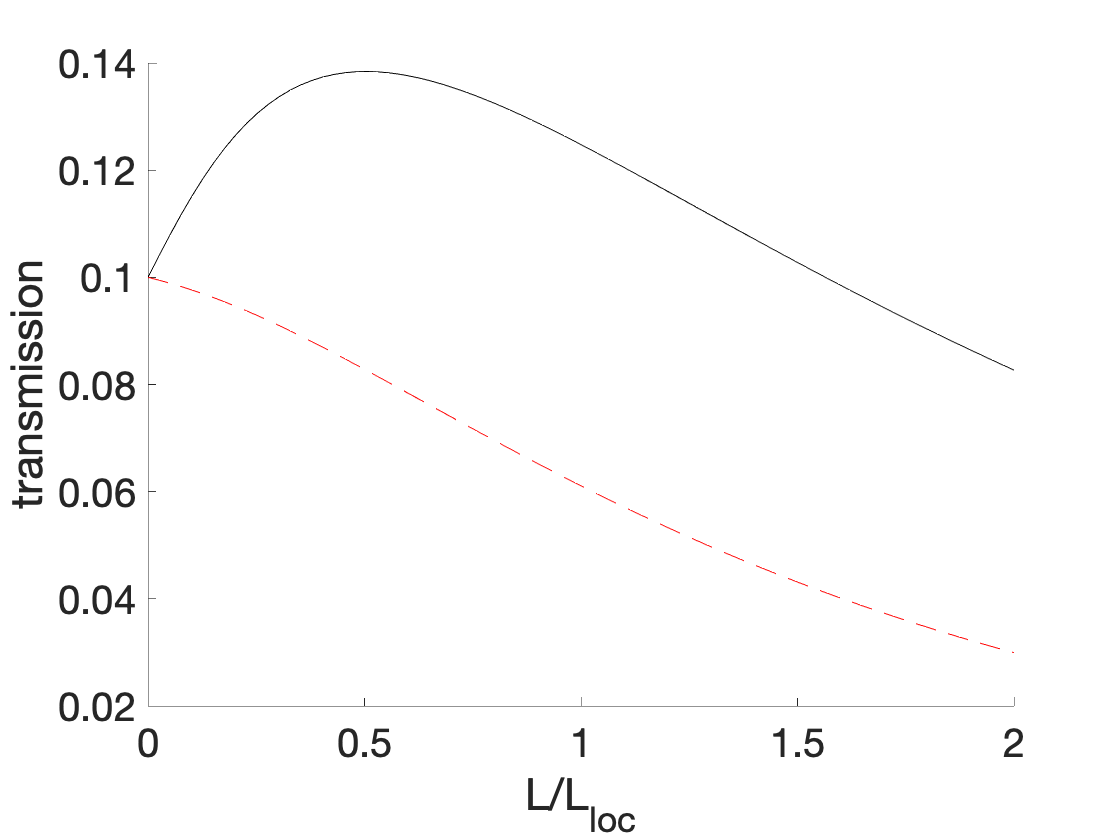} &
\includegraphics[width=2.in]{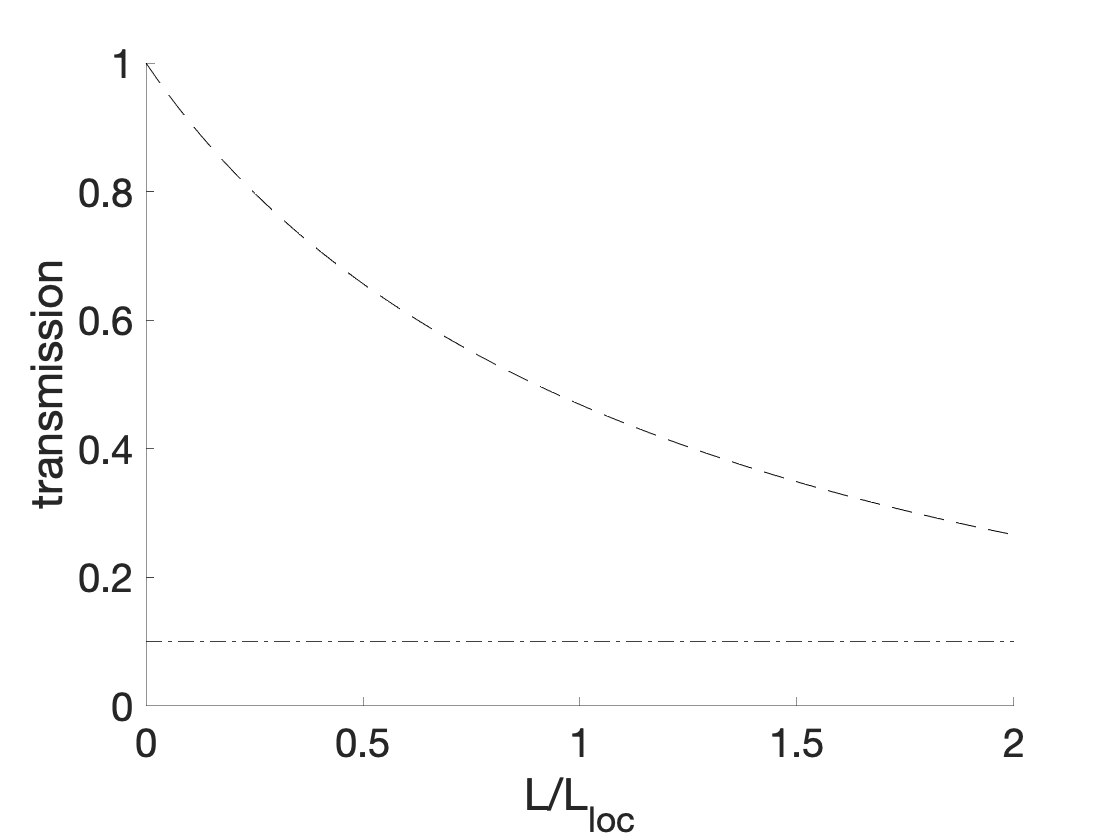}
\end{tabular}
\end{center}
\caption{Same as in Figure~\ref{fig:comp2} but but for a more reflecting barrier with  $|T_1|^2=0.1$.
}
\label{fig:comp3}
\end{figure}

Again, the transmission is enhanced by symmetry, and this is even more pronounced if the barrier is more reflecting,
as shown in Fig. \ref{fig:comp3}. See also the next case.

$\bullet$ If the barrier is strongly reflecting, i.e. $|T_1| \ll 1$, which is equivalent to having $q \gg 1$, we can use the identity 
\[
2 R_1(\om) -1 \stackrel{\eqref{eq:Reg2}}{=} \frac{q(\om)-i}{q(\om) + i} = 1 - 2 T_1(\om),
\]
in equation \eqref{eq:Tra5} to obtain 
\[
\tau_k(\om) = (k+1)^2 |T_1(\om)|^2 + o \left(|T_1(\om)|^2\right), \quad k \ge 0.
\]
Substituting this into the expression \eqref{eq:meanint} of the mean transmitted intensity we have 
\begin{equation}
\EE \left[ |\mathcal{T}(\om) |^2 \right] =|T_1(\om)|^2  \EE \left[ |T(\om)|^4 \sum_{k=0}^\infty (k+1)^2 \left(1 - |T(\om)|^2\right)^k \right].
\label{eq:mean8}
\end{equation}
This expression can be simplified using the series 
\[
\sum_{k=0}^\infty (1+k)^2 x^k = \frac{1+x}{(1-x)^3}, \qquad \forall x \in (0,1),
\]
and we obtain that 
\begin{equation}
\EE \left[ |\mathcal{T}(\om) |^2 \right] = |T_1(\om)|^2 \left\{2 \EE \left[ |T(\om)|^{-2}  \right]-1\right\} + o \left(|T_1(\om)|^2\right).
\label{eq:mean9}
\end{equation}
By solving the Kolmogorov equation 
$\partial_L U = L_{\rm loc}^{-1}
\big( 2U-1\big)$ 
satisfied by $U(L)=\EE[|T|^{-2}]$, derived using the expression of the infinitesimal generator of $|T|^2$ given in \cite[Proposition 7.3]{fouque2007wave}, 
 we get that 
\begin{equation}
2 \EE\left[|T(\om)|^{-2}\right]-1 = \exp[ 2 L/ L_{\rm loc}(\om)].
\label{eq:mean10}
\end{equation}
This result and equation \eqref{eq:mean9} show that the transmission enhancement by the random medium 
can be very large when the barrier is reflecting, as seen in Fig. \ref{fig:comp3}.

\section{Enhanced transmission in random waveguides}
\label{sect:Waveg}
In this section we study wave transmission in random 
waveguides.
To simplify the analysis, we consider two-dimensional 
waveguides with straight, sound soft boundary, as described in section \ref{sect:Waveg1}. The  mathematical model is the scalar wave equation for the pressure field. 
The decomposition of the wave into modes is  in section \ref{sect:Waveg2}. The interaction of these modes with the reflecting barrier is  in section \ref{sect:Waveg3}. The transmission and reflection of the  modes through the random sections is described in section \ref{sect:Waveg4}. The transmission through the whole system is analyzed in section 
\ref{sect:Waveg5}. We use the results in section \ref{sect:Waveg6}  to quantify the  transmission enhancement induced by symmetry, in the case of weak random scattering.

\subsection{Setup}
\label{sect:Waveg1}

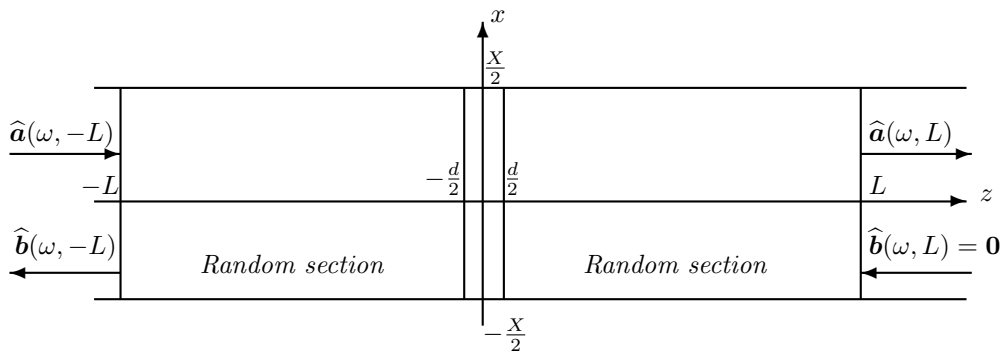
\begin{figure} 
\centerline{ 
\begin{picture}(300,75) 
\unitlength=1pt 
\thicklines 
\put(0,10){\line(1,0){330}} 
\put(0,90){\line(1,0){330}} 
\put(0,47){\vector(1,0){330}} 
\put(10,10){\line(0,1){80}} 
\put(290,10){\line(0,1){80}} 
\put(155,10){\line(0,1){80}} 
\put(140,10){\line(0,1){80}}
\put(147,0){\vector(0,1){115}}
\put(147,-6){$-\frac{X}{2}$}
\put(147,95){$\frac{X}{2}$}
\put(-5,50){$-L$} 
\put(293,50){$L$} 
\put(335,47){$z$} 
\put(125,53){$-\frac{d}{2}$} 
\put(155,53){$\frac{d}{2}$} 
\put(150,115){$x$} 
\put(40,20){{\it Random section}}
\put(185,20){{\it Random section}}
\put(10,20){\vector(-1,0){42}} 
\put(-30,27){$\hat {\itbf{b}}(\om,-L)$} 
\put(-32,65){\vector(1,0){42}} 
\put(-32,70){$\hat {\itbf a}(\om,-L)$} 
\put(332,20){\vector(-1,0){42}} 
\put(293,70){$\hat {\itbf a}(\om,L)$} 
\put(290,65){\vector(1,0){42}} 
\put(293,27){$\hat {\itbf b}(\om,L) = {\bf 0}$} 
\end{picture} 
} 
\vspace{0.1in}
\caption{Waveguide occupying the domain $\Omega = (-X/2,X/2) \times \RR$ filled at $|z| \in (d/2,L)$ with a random medium with mirror symmetry about the thin barrier located at $|z| < d/2$. }
\label{fig:4waveg}
\end{figure}
Consider a  waveguide 
occupying the domain $\Omega = (-X/2,X/2) \times \RR$ and 
introduce the system of coordinates  $\bx = (x,z)$, with $x \in (-X/2,X/2)$ and $z \in \RR$.  Assume, as illustrated in  Fig. \ref{fig:4waveg}, that the waveguide contains a thin reflecting 
barrier at $|z| < d/2$, lying between  two random sections at $|z| \in (d/2,L)$, which are mirror symmetric with respect to $z = 0$.  

The wave at frequency $\om$ is modeled by the  Fourier transform $\hat p$ of the pressure, the solution of the Helmholtz equation
\begin{equation}
\left[\frac{\om^2}{c^2(x,z)}+ \Delta\right] \hat p(\om,x,z) = 0, \qquad (x,z)  \in \Omega, 
\label{eq:Wa1}
\end{equation}
with Dirichlet boundary condition at the sound soft boundary $x = \pm X/2$,
\begin{equation}
\hat p(\om,\pm X/2,z) = 0, \qquad z \in \RR,
\label{eq:Wa2}
\end{equation}
and outgoing boundary condition at $z \to +\infty$. 
The medium that fills the waveguide is heterogeneous, with  wave speed $c$ of the form
\begin{equation}
c^{-2}(x,z) = \left\{ \begin{array}{ll} 
c_0^{-2} \qquad &\mbox{if} ~~ |z| > L ,\\
c_1^{-2} &\mbox{if} ~~ |z| <d/2 ,\\
c_0^{-2} \left[ 1 + \mu(x,|z|)\right] &\mbox{if} ~~ d/2 \le |z| \le L.
\end{array} \right.
\label{eq:Wa3}
\end{equation}
Here $c_0$ and $c_1$  are constants satisfying $c_1 < c_0$, and $\mu$ is a zero mean, mixing random process, with the uniform bound $|\mu| < 1$. 

The excitation is defined by a right going wave impinging on the 
random medium at $z = -L$ and our goal is to quantify the transmitted wave at $z = L$.

\subsection{Mode decomposition outside the barrier}
\label{sect:Waveg2}
We are interested in the case of small standard deviation of the fluctuations $\mu$ of $c^{-2}$, so we define the 
wave decomposition at $|z| >d/2$ in the reference medium with wave speed $c_0$. 

The decomposition uses the 
spectrum of the self-adjoint, negative definite operator $\partial_x^2$ with Dirichlet boundary conditions at $x = \pm X/2$. 
{The eigenvalues are given by $-\la_j$, where  $\la_j =  (j \pi/X)^2$ and the eigenfunctions are $\varphi_j(x) = \sqrt{2/X} \sin(j \pi x /X)$, for $j \ge 1$. These  form an orthonormal basis of $L^2(-X/2,X/2)$.}

{Let  $k(\om) = \om/c_0$ be the wavenumber and define the  natural number 
\begin{equation}
N(\om) = \lfloor k(\om) X/\pi \rfloor,
\label{eq:defNom}
\end{equation}
such that 
\begin{equation}
\lambda_{N(\om)} \le k^2(\om) < \lambda_{N(\om)+1}.
\label{eq:Wa4}
\end{equation}
Here $\lfloor \cdot \rfloor$ denotes the integer part. }
The wave decomposition is 
\begin{equation}
\hat p(\om,x,z) = \sum_{j=1}^{\infty} \varphi_j(x) \hat p_j(\om,z),
\label{eq:ModeDec}
\end{equation}
where $\hat p_j$ are one-dimensional, time-harmonic waves, called waveguide modes. The first $N$ of them are 
propagating waves, with  wavenumbers
\begin{equation}
\beta_j(\om) = 
\sqrt{k^2(\om) -\la_j}, \quad \mbox{if} ~ j \le N(\om),
\label{eq:Wa7}
\end{equation}
and the remaining ones are evanescent waves. These decay exponentially in $|z|$ on the length scale $\beta_j^{-1}$, where 
\begin{equation}
\beta_j(\om) = 
\sqrt{ \la_j - k^2(\om)} , \quad \mbox{if} ~ j > N(\om).
\label{eq:Wa7e}
\end{equation}
Note that if $k^2 = \lambda_{N}$, the wave $\hat p_N$ does not propagate. The analysis 
of waveguides with such standing modes is more involved than needed in this paper, so we assume 
that $\beta_N > 0$. 

The propagating waves can be decomposed further into right (forward) and left (backward) going modes, using the following equations \cite[Chapter 20]{fouque2007wave}
\begin{align}
\hat p_j(\om,z) &= \frac{1}{\sqrt{\beta_j(\om)}} \left[ \hat a_j(\om,z) e^{i \beta_j(\om)z} + 
\hat b_j(\om,z) e^{-i \beta_j(\om)z} \right], \label{eq:Wa8} \\
\partial_z \hat p_j(\om,z) &=i\sqrt{\beta_j(\om)} \left[ \hat a_j(\om,z) e^{i \beta_j(\om)z} -
\hat b_j(\om,z) e^{-i \beta_j(\om)z} \right]. \label{eq:Wa9}
\end{align}
The complex valued amplitudes of these modes are gathered in the vector fields
\begin{equation}
\hat \ba(\om,z) = \begin{pmatrix}\hat{a}_1(\om,z) \\ \vdots\\ \hat a_{N(\om)}(\om,z) \end{pmatrix}, \quad 
\hat \bb(\om,z) = \begin{pmatrix}\hat{b}_1(\om,z) \\ \vdots\\ \hat b_{N(\om)}(\om,z) \end{pmatrix},
\label{eq:Wa10}
\end{equation}
and they satisfy the  coupled system of equations 
\begin{equation}
\partial_z \begin{pmatrix} \hat \ba(\om,z) \\
\hat \bb(\om,z) \end{pmatrix} = \begin{pmatrix} \bH(\om,z) & \bK(\om,z) \\
\overline{\bK(\om,z)} & \overline{\bH(\om,z)} \end{pmatrix} \begin{pmatrix} \hat \ba(\om,z) \\
\hat \bb(\om,z) \end{pmatrix}, 
\label{eq:Wa11}
\end{equation}
derived in  \cite[Chapter 20]{fouque2007wave}. The derivation involves  substituting  \eqref{eq:ModeDec}, 
(\ref{eq:Wa8}-\ref{eq:Wa9}) into \eqref{eq:Wa1}, using the orthonormality of the 
eigenfunctions and also expressing the evanescent modes in terms of the propagating ones 
\cite[Section 20.2.3]{fouque2007wave}. The matrices $\bH, \bK \in \mathbb{C}^{N\times N}$ 
are given explicitly in \cite[Section 20.2.4]{fouque2007wave}. They depend on the mode wavenumbers 
(\ref{eq:Wa7}-\ref{eq:Wa7e}) and the  random process $\bnu = \left(\nu_{j,l}\right)_{j,l \ge 1}$, with components 
\begin{equation}
\nu_{j,l}(|z|) = \int_{-X/2}^{X/2} dx \, \varphi_j(x) \varphi_l(x) \mu(x,|z|), \qquad j,l \ge 1.
\label{eq:Wa12}
\end{equation}
In the absence of fluctuations, the matrices $\bH$ and $\bK$ would be zero i.e., 
the mode amplitudes would be decoupled and constant. This is the case at $|z| > L$, where the wave speed equals the constant $c_o$.

The system of ODEs \eqref{eq:Wa11} is complemented with the excitation $\hat \ba(\om,-L)$ 
that specifies the incoming wave impinging on the random medium and the outgoing boundary condition 
$\hat \bb(\om,L) = {\bf 0}$. Our goal is to characterize the 
transmitted mode amplitudes $\hat \ba(\om,L)$. This requires the analysis of the transmission and reflection of the 
modes at the thin barrier, described next.

\subsection{Transmission and reflection at the barrier}
\label{sect:Waveg3}
The mode decomposition inside the barrier is similar to that in equations (\ref{eq:ModeDec}-\ref{eq:Wa9}), 
except that the wave speed $c_0$ is replaced by $c_1$. Since we assume that $c_1<c_0$, we deduce from equation \eqref{eq:defNom} and its analogue inside the barrier that there are 
\begin{equation}
N_1(\om) > N(\om)
\label{eq:Wa13}
\end{equation} 
propagating modes at $|z| < d/2$. The  modes are uncoupled, with constant amplitudes, because the wave speed is constant 
inside the barrier.

The $x-$profiles of the modes inside and outside the barrier are given by the same eigenfunctions $\varphi_j$ for all $z \in \RR$, so to analyze the wave reflection and transmission  at the barrier,  it is sufficient to match $\{\hat p_j, \partial_z \hat p_j\}_{j=1}^N$ at $z = \pm d/2$. For $j \ge N+1$ the modes impinging on the barrier are evanescent and their amplitude is negligible for large enough $L$. 

The next lemma describes the propagator of the barrier. Its proof follows from the continuity of the first $N$ modes,  using a calculation that is  similar to the proof of Lemma \ref{lem.1} in Appendix \ref{ap:1DAp.1}.

\vspace{0.05in}\begin{lem}
\label{lem.Wa1}
We have 
\begin{equation}
\begin{pmatrix} 
\hat \ba(\om,d/2) \\ \hat \bb(\om,d/2) \end{pmatrix} = {\bf P}_1(\om) \begin{pmatrix} 
\hat \ba(\om,-d/2) \\ \hat \bb(\om,-d/2) \end{pmatrix}, \qquad 
{\bf P}_1(\om) = \begin{pmatrix} {\bf P}^{(a)}_1(\om) & \overline{{\bf P}^{(b)}_1(\om)} \\
{\bf P}^{(b)}_1(\om) & \overline{{\bf P}^{(a)}_1(\om)}
\end{pmatrix},
\label{eq:Wa14}
\end{equation}
where $\bf{P}_1$ is the $2N \times 2N$ propagator matrix of the barrier, with diagonal blocks 
\begin{align}
{\bf P}^{(a)}_1(\om) &= \mbox{diag} \left(\alpha_j(\om)\right)_{j = 1}^{N(\om)}  \quad \mbox{and} \quad 
{\bf P}^{(b)}_1(\om) = \mbox{diag} \left(\gamma_j(\om)\right)_{j = 1}^{N(\om)}.
\label{eq:Wa15}
\end{align}
The entries of these blocks are 
\begin{align}
\alpha_j(\om) &= \left[ \cos \Big( \beta_{1,j}(\om) d \Big) + \frac{i}{2}
\left(\frac{\beta_{1,j}(\om)}{\beta_j(\om)} +  \frac{\beta_{j}(\om)}{\beta_{1,j}(\om)}\right) 
\sin \Big( \beta_{1,j}(\om) d \Big) \right], \label{eq:Wa17} \\
\gamma_j(\om)&= \frac{i}{2}  \left(\frac{\beta_{j}(\om)}{\beta_{1,j}(\om)} -  \frac{\beta_{1,j}(\om)}{\beta_{j}(\om)}\right) 
\sin \Big( \beta_{1,j}(\om) d \Big), \label{eq:Wa18} 
\end{align}
and 
\begin{equation}
\beta_{1,j}(\om) = \sqrt{\left(\frac{\om}{c_1}\right)^2 - \la_j}, \qquad j = 1, \ldots, N(\om),
\label{eq:Wa19}
\end{equation} 
are the mode wavenumbers inside the barrier.
\end{lem} 

\vspace{0.05in}

As we have done in section \ref{sect:1D_2}, we derive from the propagator ${\bf P}_1$ the scattering matrix ${\bf S}_1
\in \mathbb{C}^{2N \times 2N}$ of the barrier. This  relates the amplitudes of the modes impinging on the barrier to those leaving the barrier,
\begin{equation}
\begin{pmatrix} 
\hat \ba(\om,d/2) \\
\hat \bb(\om,-d/2) 
\end{pmatrix} = {\bf S}_1(\om) \begin{pmatrix} 
\hat \ba(\om,-d/2) \\
\hat \bb(\om,d/2)
\end{pmatrix},
\label{eq:Wa20}
\end{equation}
and has the block structure
\begin{equation}
{\bf S}_1(\om) = \begin{pmatrix} {\bf T}_1(\om) &{\bf R}_1(\om) \\
{\bf R}_1(\om) & {\bf T}_1(\om),
\end{pmatrix}\label{eq:Wa21}
\end{equation}
with diagonal $N\times N$ blocks 
\begin{equation}
{\bf T}_1(\om) = \mbox{diag} \left(1/\overline{\alpha_j(\om)} \right)_{j=1}^{N(\om)} \quad 
\mbox{and} \quad {\bf R}_1(\om) = \mbox{diag} \left(-\gamma_j(\om)/\overline{\alpha_j(\om)} \right)_{j=1}^{N(\om)},
\label{eq:Wa22}
\end{equation}
containing the mode-dependent transmission and reflection coefficients of the barrier.

Similar to the layered case, we are interested in  the asymptotic regime 
\begin{equation}
k(\om) d \to 0, \quad \frac{c_0}{c_1} \to \infty, \quad \mbox{such that} \quad 
\left(\frac{c_0}{c_1}\right)^2 k(\om) d = O(1).
\label{eq:Wa23}
\end{equation}
In this regime, we deduce from the expressions (\ref{eq:Wa17}-\ref{eq:Wa18}) of the coefficients that 
define the propagator  that 
\begin{equation}
\alpha_j(\om) \approx 1 + i q_j(\om) \quad \mbox{and} \quad 
\gamma_j(\om) \approx -i q_j(\om),
\label{eq:Wa24}
\end{equation}
where
\begin{equation}
q_j(\om) =\frac{\beta_{1,j}^2(\om) d}{2\beta_j(\om)} \stackrel{\eqref{eq:Wa23}}{=}  O(1), \qquad j = 1, \ldots, N.
\label{eq:Wa25}
\end{equation}
The asymptotic approximation of the transmission and reflection coefficients 
is 
\begin{align}
T_{1,j}(\om) &\stackrel{\eqref{eq:Wa22}}{=} \frac{1}{\overline{\alpha_j(\om)}} \stackrel{\eqref{eq:Wa24}}{\approx} 
\frac{1}{1- iq_j(\om)},\label{eq:Wa26} \\
 \quad R_{1,j}(\om) &\stackrel{\eqref{eq:Wa22}}{=} \frac{-\gamma_j(\om)}{\overline{\alpha_j(\om)}} \stackrel{\eqref{eq:Wa24}}{\approx} 
\frac{iq_j(\om)}{1-i q_j(\om)}, \qquad j = 1, \ldots, N(\om).
\label{eq:Wa27}
\end{align}

\subsection{Transmission and reflection in the random sections}
\label{sect:Waveg4}
We collect here the relevant results from \cite[Chapter 20]{fouque2007wave} and \cite{garnier2008effective} on wave propagation in random waveguides. 
As stated at the beginning of section \ref{sect:Waveg2}, we are interested 
in small random fluctuations $\mu$ of $c^{-2}$. These have a nontrivial effect at a long distance $L$ of propagation with respect to the correlation length $\ell_c$ of the fluctuations and the wavelength $\la$. 
Thus, we consider the asymptotic regime 
\begin{equation}
\ell_c \sim \lambda \sim X \ll L, \qquad {\rm Var}(\mu)\ll 1,
\label{eq:regime2}
\end{equation}
where we deduce from equation \eqref{eq:defNom} that the number $N$ of propagative modes is of order one. The scattering effect of the random medium  on the transmittivity is of order one when ${\rm Var}(\mu) \ell_c L / \lambda^2 = O(1)$  and it is smaller than one when ${\rm Var}(\mu) \ell_c L \ll \lambda^2$. The latter defines what we call the weak scattering regime and is of particular interest in this paper because it allows the explicit quantification of the mean transmittivity of the waveguide
(see section \ref{sect:Waveg6}).

The propagator matrix ${\bf P}_{-}$ for the left random section is
the solution of 
\begin{align}
\partial_z {\bf P}_{-}(\om,z) &=  \begin{pmatrix} \bH(\om,z) & \bK(\om,z) \\ 
\overline{\bK(\om,z)} & \overline{\bH(\om,z)} \end{pmatrix} {\bf P}_{-}(\om,z), \qquad z \in (-L, -d/2), 
\label{eq:Wa29} \\
{\bf P}_{-}(\om,-d/2) &= \bI_{2N} \label{eq:Wa30}
\end{align}
where $\bI_{2N}$ denotes the $2N \times 2N$ identity matrix. Given the algebraic form of the coupling matrix in the right hand side of \eqref{eq:Wa29}, one can deduce that the propagator has the block form \cite[Section 20.2.5]{fouque2007wave}
\begin{equation}
{\bf P}_{-}(\om,z) = 
\begin{pmatrix} {\bf P}_-^{(a)}(\om,z) &\overline{{\bf P}_-^{(b)}(\om,z)} \\
{\bf P}_-^{(b)}(\om,z) & \overline{{\bf P}_-^{(a)}(\om,z)}
\end{pmatrix},\label{eq:Wa31}
\end{equation}
with full blocks ${\bf P}_-^{(a)}, {\bf P}_-^{(b)} \in \mathbb{C}^{N \times N}$  that capture mode coupling induced by scattering in the random medium. 
We are interested in the propagator evaluated at $z = -L$, which defines the $N \times N$ 
transmission and reflection matrices 
of the left random section. These matrices are the analogues of the scalar valued transmission and reflection coefficients in layered media, deduced from the propagator as explained in Appendix \ref{ap:1DAp.2}. We have 
\begin{equation}
\begin{pmatrix} \bI\\ 
{\bf R}_-(\om) \end{pmatrix} = {\bf P}_{-}(\om,-L) \begin{pmatrix} {\bf T}_-(\om) \\ 
{\bf 0} \end{pmatrix},
\label{eq:Wa32}
\end{equation}
which can be understood from the waveguide analogue of Fig. \ref{fig:reftr1} and 
\begin{equation}
\begin{pmatrix} {\bf 0} \\ 
\tilde {\bf T}_-(\om) \end{pmatrix} = {\bf P}_{-}(\om,-L) \begin{pmatrix}\tilde  {\bf R}_-(\om) \\ 
\bI \end{pmatrix},
\label{eq:Wa33}
\end{equation}
which corresponds to the analogue of Fig. \ref{fig:reftr2}. Here ${\bf 0}$ and $\bI$ are the $N \times N$ zero and identity matrices, respectively. 

Similarly, the propagator ${\bf P}_{+}$ for the right random section is the solution of
\begin{align}
\partial_z {\bf P}_{+}(\om,z) &=  \begin{pmatrix} \bH(\om,z) & \bK(\om,z) \\ 
\overline{\bK(\om,z)} & \overline{\bH(\om,z)} \end{pmatrix} {\bf P}_{+}(\om,z), \qquad z \in (d/2,L), 
\label{eq:Wa34} \\
{\bf P}_{+}(\om,d/2) &= \bI_{2N}, \label{eq:Wa35}
\end{align}
and its algebraic structure is like in equation \eqref{eq:Wa31}, with $N \times N$ blocks 
${\bf P}^{(a)}_+$ and ${\bf P}^{(b)}_+$.
This propagator defines the  $N \times N$ transmission and reflection matrices of the right random section according to equations 
\begin{equation}
\begin{pmatrix} 
{\bf T}_+(\om)  \\{\bf 0}\end{pmatrix} = {\bf P}_{+}(\om,L) \begin{pmatrix}{\bf I}\\  {\bf R}_+(\om) \end{pmatrix},
\label{eq:Wa36}
\end{equation}
and 
\begin{equation}
\begin{pmatrix} 
\tilde {\bf R}_+(\om) \\ {\bf I}\end{pmatrix} = {\bf P}_{+}(\om,L) \begin{pmatrix}{\bf 0} \\ \tilde  {\bf T}_+(\om)  \end{pmatrix}.
\label{eq:Wa37}
\end{equation}
These can be understood from the waveguide analogues of Fig. \ref{fig:reftr3}-\ref{fig:reftr4}. 

Note the symmetry of the definitions (\ref{eq:Wa29}-\ref{eq:Wa30}) and (\ref{eq:Wa34}-\ref{eq:Wa35}). Both propagators start as the identity $\bI_{2N}$ at $z = \pm d/2$ and 
define the transmission and reflection matrices at $z = \pm L$. The expression of the coupling matrices $\bH$ and $\bK$
given in \cite[Section 20.2.4]{fouque2007wave} and the symmetry of the fluctuations about $z = 0$, give that 
\begin{equation}
\bH(\om,z) = - \overline{\bH(\om,-z)} \quad \mbox{and} \quad 
\bK(\om,z) = - \overline{\bK(\om,-z)}.
\label{eq:Wa38}
\end{equation}
This implies that 
\begin{equation}
{\bf P}_-(\om,-L) = \overline{{\bf P}_+(\om,L)},
\label{eq:Wa39}
\end{equation}
and solving equations (\ref{eq:Wa32}-\ref{eq:Wa33}) and 
(\ref{eq:Wa36}-\ref{eq:Wa37}), we get:  The transmission matrices satisfy 
\begin{align}
{\bf T}_+(\om) &= 
{\tilde {\bf T}_-(\om) } = {\bf P}^{(a)}_{+}(\om,L) - 
\overline{{\bf P}^{(b)}_{+}(\om,L)} \left[ \overline{{\bf P}^{(a)}_{+}(\om,L)}\right]^{-1} 
{\bf P}^{(b)}_{+}(\om,L) ,\nonumber \\
\tilde {\bf T}_+(\om) &= { {\bf T}_-(\om) } = \left[ \overline{{\bf P}^{(a)}_{+}(\om,L)}\right]^{-1},
\label{eq:Wa40}
\end{align}
and the reflection matrices satisfy 
\begin{align}
{\bf R}_+(\om) &={\tilde {\bf R}_-(\om) } =  -\left[ \overline{{\bf P}^{(a)}_{+}(\om,L)}\right]^{-1}
{\bf P}^{(b)}_{+}(\om,L),\nonumber \\
\tilde {\bf R}_+(\om) &= { {\bf R}_-(\om) } = \overline{{\bf P}^{(b)}_{+}(\om,L)} \left[ \overline{{\bf P}^{(a)}_{+}(\om,L)}\right]^{-1} .
\label{eq:Wa41}
\end{align}
In addition, we have the energy conservation relation 
\cite[Eq. (20.41)]{fouque2007wave}
\begin{equation}
{\bf R}^{\star}_+(\om) {\bf R}_+(\om) + {\bf T}^{\star}_+(\om) {\bf T}_+(\om) = \bI,
\label{eq:Wa42}
\end{equation} 
and the reciprocity relations \cite[Page 1582]{garnier2008effective}
\begin{equation}
{\bf R}^{T}_+(\om) \approx {\bf R}_+(\om) \quad \mbox{and} \quad 
\tilde {\bf R}^{T}_+(\om) \approx \tilde {\bf R}_+(\om).
\label{eq:Wa43}
\end{equation} 
Here the index $T$ stands for transpose, the star $\star$ denotes the complex conjugate and transpose and the approximation 
in \eqref{eq:Wa43} means that reciprocity holds in the asymptotic regime \eqref{eq:regime2}. 
\subsection{Transmission through the system}
\label{sect:Waveg5}
The propagator matrix $\cbP$ for the waveguide is defined by the equation
\begin{equation}
\begin{pmatrix} 
\hat \ba(\om,L) \\
\hat \bb(\om,L)
\end{pmatrix} = \cP(\om) \begin{pmatrix} 
\hat \ba(\om,-L) \\
\hat \bb(\om,-L)
\end{pmatrix}.
\label{eq:Wa46}
\end{equation}
From the definitions \eqref{eq:Wa14}, \eqref{eq:Wa29} and \eqref{eq:Wa34} of the propagators of the barrier and the random sections, and the identity \eqref{eq:Wa39},   we deduce that 
\begin{equation}
\cbP(\om) =  {\bf P}_+(\om,L) {\bf P}_1(\om) \left[ \overline{{\bf P}_+(\om,L)}\right]^{-1}.
\label{eq:Wa47}
\end{equation}
Recalling that $\hat \bb(\om,L) = {\bf 0}$ and that the excitation
specifies the incoming mode amplitudes stored in  $\hat \ba(\om,-L)$, we can define the transmission and reflection matrices 
$\cbT, \cbR \in \mathbb{C}^{N \times N}$ of the system by the following equation
\begin{equation}
\begin{pmatrix} 
\bI \\ \cbR(\om) 
\end{pmatrix} = \cbP(\om)
\begin{pmatrix} 
\cbT(\om) \\{\bf 0}
\end{pmatrix}.
\label{eq:Wa48}
\end{equation} 
The expression of the transmission matrix $\cbT$ is given in the next theorem, proved in Appendix \ref{ap:WavegAp.1}.

\vspace{0.05in} 
\begin{thm}
\label{thm.W1}
The $N \times N$ transmission matrix for the waveguide has the expression
\begin{align}
\cbT(\om) \approx {\bf T}_+(\om)  \left[ {\bf T}_1^{-1}(\om) - 
{\bf R}_+(\om) {\bf T}_1^{-1}(\om){\bf R}_1(\om) -{\bf T}_1^{-1}(\om){\bf R}_1(\om){\bf R}_+(\om) \right. \\
\left. -{\bf R}_+(\om) \overline{{\bf T}_1^{-1}(\om)} {\bf R}_+(\om) \right]^{-1}{\bf T}^{T}_+(\om),
\label{eq:Wa49}
\end{align}
where  the approximation holds in the regime \eqref{eq:regime2}.
\end{thm}

\vspace{0.05in} The transmissivity of the system is 
\begin{equation}
\mbox{Tr} \left[  {\cbT}^{\star}(\om) {\cbT}(\om) \right] = 
\sum_{j,l = 1}^{N(\om)} \left| \cT_{jl}(\om)\right|^2,
\label{eq:Wa50}
\end{equation} 
where ``Tr" denotes the trace. In the next section we quantify the mean of \eqref{eq:Wa50} in the asymptotic 
regime \eqref{eq:regime2}.

\subsection{Enhanced transmission}
\label{sect:Waveg6}
To quantify the effect of symmetry on the wave transmission through the waveguide, we derive next the expression of the mean transmissivity. This requires the statistical moments of the products of the entries of the transmission and reflection matrices ${\bf T}_+$ and ${\bf R}_+$. These moments are characterized in the regime \eqref{eq:regime2}  in 
\cite[Propositions 3.1, 4.2]{garnier2008effective}. Their expression is very complicated, so we do not repeat it here.
However, the result simplifies in the case of weak scattering in the random medium: 

\vspace{0.05in} 
\begin{thm}
\label{thm.W2}
When the random medium is weakly scattering i.e., in the asymptotic regime \eqref{eq:regime2} with ${\rm Var}(\mu) \ell_c L \ll \lambda^2$,
 the mean transmissivity is approximated by 
\begin{equation}
\EE\left[ \sum_{j,l = 1}^{N(\om)} \left| \cT_{jl}(\om)\right|^2\right] \approx \TT(\om)= \sum_{l=1}^{N(\om)} \left[ \big|T_{1,l}(\om)\big|^2 + 
 \sum_{m=1}^{N(\om)} {\cal M}_{lm}(\om) {\cal B}_{lm}(\om) \right],
\label{eq:Ewa2}
\end{equation}
where we introduced the moments 
\begin{equation}
{\cal M}_{lm}(\om)  =\EE [ |R_{+,lm}(\om)|^2], 
\label{eq:Ewa3} 
\end{equation}
and the factors 
\begin{equation}
{\cal B}_{lm}(\om) = |T_{1,l}(\om) + T_{1,m}(\om) - 2 T_{1,l}(\om) T_{1,m}(\om) |^2 - |T_{1,l}(\om)|^2 - |T_{1,m}(\om)|^2,
\label{eq:Ewa4}
\end{equation}
that depend only on the barrier.
 \end{thm}

\vspace{0.05in}  
The proof of this theorem is in Appendix \ref{ap:WavegAp.2}. We conclude from its statement  that if there is no random medium, 
the transmissivity equals that of the barrier, denoted by 
\begin{equation}
\TT_0(\om) = \sum_{l=1}^{N(\om)}  \big|T_{1,l}(\om)\big|^2.
\end{equation}
If the random medium is present, its effect on the mean transmissivity depends on the strength of the barrier,  
which determines the sign of the factors \eqref{eq:Ewa4}. The moments ${\cal M}_{lm}$ are positive by definition, so if the factors 
${\cal B}_{lm}$ are positive, we have transmission enhancement induced by the symmetry of the random medium. 

Let us write more explicitly equation \eqref{eq:Ewa4},
\begin{align*} 
{\cal B}_{lm}(\om) &= 4 |T_{1,l}(\om)|^2 |T_{1,m}(\om)|^2  - 
4 |T_{1,m}(\om)|^2 \mbox{Re}[T_{1,l}(\om)] - 
4 |T_{1,l}(\om)|^2  \mbox{Re}[T_{1,m}(\om)]\\&+ 2 \mbox{Re} [ T_{1,m}(\om) \overline{T_{1,l}(\om)}],
\end{align*}
and observe from equation \eqref{eq:Wa26} that $\mbox{Re}(T_{1,l}) = |T_{1,l}|^2$. This gives that 
\begin{align}
{\cal B}_{lm}(\om) &= - 4 |T_{1,l}(\om)|^2 |T_{1,m}(\om)|^2 + 2 \mbox{Re} [ T_{1,m}(\om) \overline{T_{1,l}(\om)}] \nonumber \\
&\stackrel{\eqref{eq:Wa26}}{=} \frac{-2[1-  q_l(\om) q_m(\om)]}{[1+q_l^2(\om)][1+q_m^2(\om)]}, \qquad l, m = 1, \ldots, N(\om).
\label{eq:Ewa5}
\end{align}
Consequently, ${\cal B}_{lm} < 0$ if the barrier is weak i.e., the parameters $\{q_l\}_{l=1}^N$ are small, and  the  random medium has a negative effect on the transmissitivity, because 
\begin{equation}
\TT(\om) - \TT_0(\om) <0. 
\label{eq:Ewa6}
\end{equation}
However, if the barrier is strong enough to make the parameters   $\{q_l\}_{l=1}^N$ larger than $1$, the factors \eqref{eq:Ewa5} are positive and we have transmission enhancement 
\begin{equation}
\TT(\om) - \TT_0(\om) \approx \sum_{l=1}^{N(\om)} \sum_{m=1}^{N(\om)} {\cal M}_{lm}(\om) {\cal B}_{lm}(\om) > 0.
\label{eq:Ewa7}
\end{equation}
The enhancement is due to the symmetry of the random medium about the strong barrier. Without the symmetry, 
the mean transmissivity is reduced, as stated in the next proposition, proved in Appendix \ref{ap:WavegAp.3}.

\vspace{0.05in} 
\begin{prop}
\label{prop.W1}
When the random medium is weakly scattering i.e., in the asymptotic regime \eqref{eq:regime2} with ${\rm Var}(\mu) \ell_c L \ll \lambda^2$, and the random media in the left and right sections of the waveguide are statistically independent, the mean transmissivity of the system is approximated by 
\begin{align}
\EE\left[ \sum_{j,l = 1}^{N(\om)} \left| \cT_{jl}(\om)\right|^2\right] \approx \TT_0(\om) - 2 
\sum_{l=1}^{N(\om)} \sum_{m = 1}^{N(\om)} {\cal M}_{lm}(\om) \left|T_{1,l}(\om)\right|^2  \left|T_{1,m}(\om)\right|^2,
\label{eq:Ewa8}
\end{align} 
and is therefore smaller than the transmissivity $\mathbb{T}_0(\om)$ of the barrier.
\end{prop}

\section{Summary}
\label{sect:sum}
We have  introduced a detailed mathematical analysis of wave transmission enhancement in random systems 
with symmetry about a reflecting barrier. The analysis is motivated by recent experimental results reported in the physics literature, which observe such enhancement in symmetric cavities and in diffusive slabs.  We consider 
acoustic waves for simplicity, although the methodology applies to any linear waves. The main result is the quantification of  the mean transmissivity of two  random systems  with a preferred direction of propagation: plane waves in randomly layered media and waves in random waveguides. 
The first case is easier to analyze and we consider both weak and strongly scattering random media. The waveguide setting is 
significantly more complex, so we quantify the transmission enhancement only in the case of weakly scattering random media.
The transmission enhancement induced by symmetry is shown in both settings and it is much more pronounced for large 
opacity of the barrier.

\section* {Acknowledgements}  This material is based upon research supported in part by the  AFOSR award number FA9550-22-1-0077 and was also partially supported by Agence de l'Innovation de D\'efense - AID - via Centre Interdisciplinaire d'Etudes pour la D\'efense et la S\'ecurit\'e - CIEDS - (project PRODIPO).
We are grateful for the opportunity to participate in the 2023 program ``Mathematical theory and applications of multiple wave scattering" 
at the Isaac Newton Institute for Mathematical Sciences in Cambridge, UK, where we learned about the experiments that motivated this work.

\appendix
\section{Derivation of the results for randomly layered media}
\label{ap:1DAp}
In this appendix we prove the results stated in section \ref{sect:1D}. Since the frequency $\om$  is fixed in the 
proofs, we simplify the notation throughout the appendix, by dropping the argument $\om$ of the propagator and scattering matrices below.
\subsection{Proof of Lemma \ref{lem.1}}
\label{ap:1DAp.1}
The statement of the lemma is derived from the continuity of the Fourier coefficients of the pressure and velocity fields.
The decomposition of these fields is given  in equations (\ref{eq:modedec1}-\ref{eq:modedec2}) outside the barrier 
and their analogues inside the barrier. The medium inside the barrier is homogenenous, so it follows from  equation \eqref{eq:1D.1}  that 
the right and left going mode amplitudes there,  denoted by $\hat a_1$ and $\hat b_1$, satisfy 
\begin{equation}
\partial_z \hat a_1( z) = \partial_z \hat b_1( z) = 0, \qquad z \in (-d/2,d/2).
\label{eq:A1_1}
\end{equation}

When imposing the  continuity of the wave field at $z = -d/2$, we obtain that 
\begin{equation}
\begin{pmatrix}
\hat a_1\big( -\frac{d}{2}\big) e^{-i \om \frac{d}{2 c_1}} \\
\hat b_1\big( -\frac{d}{2} \big) e^{i \om \frac{d}{2 c_1}}
\end{pmatrix} = \begin{pmatrix} r_+ & r_- \\
r_{-} & r_+ 
\end{pmatrix} \begin{pmatrix}
\hat a\big( -\frac{d}{2}\big) e^{-i \om \frac{d}{2 c_0}} \\
\hat b\big( -\frac{d}{2} \big) e^{i \om \frac{d}{2 c_0}}
\end{pmatrix},
\label{eq:A1}
\end{equation}
where
\begin{equation}
\label{eq:rpm}
r_\pm = \frac{1}{2} \left( \sqrt{\frac{\zeta_1}{\zeta_0}} - \sqrt{\frac{\zeta_0}{\zeta_1}}\right).
\end{equation}
 The continuity at $z = d/2$ gives 
\begin{equation}
\begin{pmatrix}
\hat a\big( \frac{d}{2}\big) e^{i \om \frac{d}{2 c_0}} \\
\hat b\big( \frac{d}{2} \big) e^{-i \om \frac{d}{2 c_0}}
\end{pmatrix} = \begin{pmatrix} r_+ & -r_- \\
-r_{-} & r_+ 
\end{pmatrix} \begin{pmatrix}
\hat a_1\big( \frac{d}{2}\big) e^{i \om \frac{d}{2 c_1}} \\
\hat b_1\big( \frac{d}{2} \big) e^{-i \om \frac{d}{2 c_1}}
\end{pmatrix},
\label{eq:A2}
\end{equation}
and from equation \eqref{eq:A1_1} we have 
\begin{equation}
\hat a_1\Big(\frac{d}{2}\Big) = \hat a_1\Big(-\frac{d}{2}\Big), \quad 
\hat b_1\Big(\frac{d}{2}\Big) = \hat b_1\Big(-\frac{d}{2}\Big).
\label{eq:A3}
\end{equation}
Combining these equations we obtain 
\begin{equation}
\begin{pmatrix}
\hat a\big(\frac{d}{2}\big) \\
\hat b\big(\frac{d}{2} \big) 
\end{pmatrix} = {\bf P}_1 \begin{pmatrix}
\hat a\big(-\frac{d}{2}\big) \\
\hat b\big(-\frac{d}{2} \big) 
\end{pmatrix},
\end{equation}
where 
\begin{align}
{\bf P}_1= \
\begin{pmatrix}  e^{ -i \omega \frac{d}{2c_0}} &0\\
0& e^{ i \omega \frac{d}{2c_0}}
\end{pmatrix}
\begin{pmatrix} r_+ & -r_- \\
-r_{-} & r_+ 
\end{pmatrix}  
\begin{pmatrix}  e^{ i \omega \frac{d}{c_1}} &0\\
0& e^{ -i \omega \frac{d}{ c_1}}
\end{pmatrix} \nonumber \\
\times  \begin{pmatrix} r_+ & r_- \\
r_{-} & r_+ 
\end{pmatrix}
\begin{pmatrix}  e^{ -i \omega \frac{d}{2c_0} }&0\\
0& e^{ i \omega \frac{d}{2c_0}}
\end{pmatrix}.
\label{eq:A7}
\end{align}
Multiplying the matrices in \eqref{eq:A7} we get the algebraic form 
\eqref{eq:defPb} of ${\bf P}_1$, with 
\begin{align}
\alpha&= \left[ (r_+^2 - r_{-}^2)\cos \Big(\frac{\omega d}{c_1} \Big) + i (r_+^2 + r_{-}^2)\sin \Big(\frac{\omega d}{c_1} \Big)
\right] e^{-i \om d/c_0}, \\
\gamma &= -2 i r_+ r_{-} \sin \Big(\frac{\omega d}{c_1} \Big).
\end{align}
Finally, definition \eqref{eq:rpm} gives
\begin{equation}
r_+^2 - r_{-}^2 = 1 \quad \mbox{and} \quad 
r_+ r_- = \frac{1}{4} \left(\frac{\zeta_1}{\zeta_0}-\frac{\zeta_0}{\zeta_1} \right),
\end{equation}
and the statement of Lemma \ref{lem.1} follows. $\Box$
\subsection{Proof of Lemma \ref{lem.2}}
\label{ap:1DAp.2}
\begin{figure} 
\centerline{ 
\begin{picture}(226,75) 
\unitlength=1pt 
\thicklines 
\put(0,10){\vector(1,0){235}} 
\put(48,0){\line(0,1){80}} 
\put(193,0){\line(0,1){80}} 
\put(50,0){$d/2$} 
\put(195,0){$L$} 
\put(230,1){$z$} 
\put(85,45){{\it Random section}}
\put(42,34){\vector(-1,0){42}} 
\put(20,41){$R_+$} 
\put(0,64){\vector(1,0){42}} 
\put(20,69){$1$} 
\put(239,64){\vector(-1,0){42}} 
\put(215,69){$0$} 
\put(197,34){\vector(1,0){42}} 
\put(212,41){$T_+$} 
\end{picture} 
} 
\caption{Reflection and transmission coefficients $R_+$ and $T_+$ for the random section $(d/2,L)$.} 
\label{fig:reftr3} 
\end{figure} 
Consider first the  random section $[d/2,L]$  and define the propagator ${\bf P}_+$ of the subsection $[d/2,z]$ by  
\begin{equation}
\begin{pmatrix}
\hat a(z) \\
\hat b(z) 
\end{pmatrix} = {\bf P}_+(z) \begin{pmatrix}
\hat a\big(\frac{d}{2}\big) \\
\hat b \big(\frac{d}{2}\big)
\end{pmatrix}, \qquad z \in \Big(\frac{d}{2}, L \Big].
\label{eq:A11}
\end{equation}
It is shown in  \cite[Chapter 7 and Section 4.4.3]{fouque2007wave} that 
\begin{equation}
{\bf P}_+ (z)= \begin{pmatrix}
\alpha_+(z) & \overline{\gamma_+(z)}\\
\gamma_+(z) & \overline{\alpha_+(z)}
\end{pmatrix},
\label{eq:A12}
\end{equation}
where $\alpha_+$ and $\gamma_+$ satisfy the first order system 
\begin{equation}
\frac{d}{dz}\begin{pmatrix}
\alpha_+(z)\\
\gamma_+(z)
\end{pmatrix}
=
 \frac{i\omega}{2c_0} \mu(z)
\begin{pmatrix}1 & - e^{-2i \omega z/c_0} \\
  e^{2i \omega z/c_0} &-1
\end{pmatrix}
\begin{pmatrix}
\alpha_+(z)\\
\gamma_+(z)
\end{pmatrix},
\label{eq:A13}
\end{equation}
at $z \in (d/2,L)$, and the initial conditions
\begin{equation} 
\alpha_+ \Big(\frac{d}{2} \Big) = 1, \quad {\gamma_+\Big(\frac{d}{2} \Big) = 0.}
\label{eq:A14}
\end{equation}
This is illustrated schematically in Fig. \ref{fig:reftr3} and at $z = L$ we have 
\begin{equation}
\begin{pmatrix} 
T_+ \\ 0 \end{pmatrix} = {\bf P}_+(L) \begin{pmatrix} 
1 \\ R_+ 
\end{pmatrix},
\label{eq:A15}
\end{equation}
where $T_+$ and $R_+$ are the random transmission and reflection coefficients, defined by  
\begin{equation}
T_+ = \frac{1}{\overline{\alpha_+(L)}}, \quad 
R_+ = - \frac{\gamma_+(L)}{\overline{\alpha_+(L)}}.
\label{eq:A16}
\end{equation}

\begin{figure} 
 \centerline{ 
\begin{picture}(226,75) 
\unitlength=1pt 
\thicklines 
\put(0,10){\vector(1,0){235}} 
\put(48,0){\line(0,1){80}} 
\put(193,0){\line(0,1){80}} 
\put(50,0){$d/2$} 
\put(195,0){$L$} 
\put(230,1){$z$} 
\put(85,45){{\it Random section}}
\put(42,34){\vector(-1,0){42}} 
\put(20,41){$\tilde{T}_+ $} 
\put(0,64){\vector(1,0){42}} 
\put(20,69){$0$} 
\put(239,64){\vector(-1,0){42}} 
\put(219,69){$1$} 
\put(197,34){\vector(1,0){42}} 
\put(215,41){$\tilde{R}_+$} 
\end{picture} 
} 
\caption{Adjoint reflection and transmission coefficients $\tilde{R}_+$ and $\tilde{T}_+$ for  $z \in (d/2,L)$. } 
\label{fig:reftr4} 
\end{figure}
Since  the matrix in equation \eqref{eq:A13} has trace zero, we have the conservation relation  \cite[Section 7.1.1]{fouque2007wave} 
\begin{equation}
\mbox{det} \left[{\bf P}_+(L)\right] = |\alpha_+(L)|^2 - |\gamma_+(L)|^2 = 1,
\label{eq:A17}
\end{equation}
which in light of definitions \eqref{eq:A16} is equivalent to $
|R_+|^2 + |T_+|^2 = 1.$ 
Because of this relation, the inverse of the propagator is
\begin{equation}
{\bf P}_+^{-1}(L) = \begin{pmatrix} 
\overline{\alpha_+(L)} & - \overline{\gamma_+(L)} \\
-\gamma_+(L) & \alpha_+(L)
\end{pmatrix}, 
\label{eq:A19}
\end{equation}
and from \eqref{eq:A15} we obtain that  
\begin{equation}
\begin{pmatrix} 
1 \\ R_+
\end{pmatrix} = \begin{pmatrix} 
\overline{\alpha_+(L)} & - \overline{\gamma_+(L)} \\
-\gamma_+(L) & \alpha_+(L) 
\end{pmatrix} \begin{pmatrix} 
T_+ \\ 0 \end{pmatrix}.
\label{eq:A20}
\end{equation}
Reordering these equations and defining 
\begin{equation}
\tilde T_+ = T_+ = \frac{1}{\overline{\alpha_+(L)}}, \quad \tilde R_+ = \frac{\overline{\gamma_+(L)}}{\overline{\alpha_+(L)}},
\label{eq:A21}
\end{equation}
we obtain the adjoint problem, illustrated schematically in Fig. \ref{fig:reftr4} ,
\begin{equation}
\begin{pmatrix} 
\tilde R_+ \\
1 
\end{pmatrix} = {\bf P}_+(L) \begin{pmatrix} 
0 \\
\tilde{T}_+(L) 
\end{pmatrix}.
\label{eq:A22}
\end{equation}

Now we can obtain from equation \eqref{eq:A11} evaluated at $z = L$ and the definitions \eqref{eq:A16} and 
\eqref{eq:A21} of the transmission and reflection coefficients that 
\begin{equation}
\begin{pmatrix} \hat a(L) \\
\hat b\big(\frac{d}{2} \big) 
\end{pmatrix} = \underbrace{\begin{pmatrix} 
T_+ & \tilde R_+ \\
R_+ & T_+ 
\end{pmatrix}}_{{\bf S}_+} \begin{pmatrix} \hat a \big(\frac{d}{2} \big) \\
\hat b(L) 
\end{pmatrix},
\label{eq:A23}
\end{equation}
where ${\bf S}_+$ is the scattering matrix of the random section $[d/2,L]$.

Similarly, the propagator matrix for the left random section satisfies 
\begin{equation}
\begin{pmatrix}
\hat a(z) \\
\hat b(z) 
\end{pmatrix} = {\bf P}_-(z) \begin{pmatrix}
\hat a\big(-\frac{d}{2}\big) \\
\hat b \big(-\frac{d}{2}\big)
\end{pmatrix}, \qquad z \in \Big[-L,-\frac{d}{2} \Big),
\label{eq:24}
\end{equation}
where 
\begin{equation}
{\bf P}_- (z)= \begin{pmatrix}
\alpha_-(z) & \overline{\gamma_-(z)}\\
\gamma_-(z) & \overline{\alpha_-(z)}
\end{pmatrix},
\label{eq:A25}
\end{equation}
and $\alpha_-$ and $\beta_-$ satisfy 
\begin{equation}
\frac{d}{dz}\begin{pmatrix}
\alpha_-(z)\\
\gamma_-(z)
\end{pmatrix}
=
 \frac{i\omega}{2c_0} \mu(-z)
\begin{pmatrix}1 & - e^{-2i \omega z/c_0} \\
  e^{2i \omega z/c_0} &-1
\end{pmatrix}
\begin{pmatrix}
\alpha_-(z)\\
\gamma_-(z)
\end{pmatrix}, 
\label{eq:A26}
\end{equation}
at $z \in (-L,-d/2)$, and the initial  conditions
\begin{equation} 
\alpha_- \Big(-\frac{d}{2} \Big) = 1, \quad \gamma_- \Big(-\frac{d}{2} \Big) = 0.
\label{eq:A27}
\end{equation}
Note that due to the symmetry of the random medium, $\left( \overline{\alpha_-(-z)},\overline{\gamma_-(-z)}\right)$ satisfies the same equation and initial condition as $ \left({\alpha_+(z)},{\gamma_+(z)}\right)$. Therefore,
\begin{equation}
\alpha_{-}(-L) = \overline{\alpha_+(L)}, \quad 
\gamma_{-}(-L) = \overline{\gamma_+(L)}.
\label{eq:A28}
\end{equation}

\begin{figure} 
\centerline{ 
\begin{picture}(226,75) 
\unitlength=1pt 
\thicklines 
\put(0,10){\vector(1,0){235}} 
\put(48,0){\line(0,1){80}} 
\put(193,0){\line(0,1){80}} 
\put(50,0){$-L$} 
\put(195,0){$-d/2$} 
\put(230,1){$z$} 
\put(85,45){{\it Random section}}
\put(42,34){\vector(-1,0){42}} 
\put(20,41){$R_-$} 
\put(0,64){\vector(1,0){42}} 
\put(20,69){$1$} 
\put(239,64){\vector(-1,0){42}} 
\put(215,69){$0$} 
\put(197,34){\vector(1,0){42}} 
\put(215,41){$T_-$} 
\end{picture} 
} 
\caption{Reflection and transmission coefficients $R_-$ and $T_-$ for random section $(-L,-d/2)$. } 
\label{fig:reftr1} 
\end{figure} 
 
 \begin{figure} 
 \centerline{ 
\begin{picture}(226,75) 
\unitlength=1pt 
\thicklines 
\put(0,10){\vector(1,0){235}} 
\put(48,0){\line(0,1){80}} 
\put(193,0){\line(0,1){80}} 
\put(50,0){$-L$} 
\put(195,0){$-d/2$} 
\put(230,1){$z$} 
\put(85,45){{\it Random section}}
\put(42,34){\vector(-1,0){42}} 
\put(20,41){$\tilde{T}_- $} 
\put(0,64){\vector(1,0){42}} 
\put(20,69){$0$} 
\put(239,64){\vector(-1,0){42}} 
\put(219,69){$1$} 
\put(197,34){\vector(1,0){42}} 
\put(215,41){$\tilde{R}_-$} 
\end{picture} 
} 
\caption{Adjoint reflection and transmission coefficients $\tilde{R}_-$ and $\tilde{T}_-$ for $z \in (-L,-d/2)$.} 
\label{fig:reftr2} 
\end{figure} 
The reflection and transmission through the left random section is illustrated schematically in Figs.~\ref{fig:reftr1}-\ref{fig:reftr2} and the transmission and reflection coefficients are defined by 
\begin{equation*}
\begin{pmatrix} 1 \\
R_- \end{pmatrix} = {\bf P}_-(-L) \begin{pmatrix} T_- \\
0 \end{pmatrix} \quad \mbox{and} \quad 
\begin{pmatrix} 0 \\
\tilde{T}_- \end{pmatrix} = {\bf P}_-(-L) \begin{pmatrix} \tilde R_- \\
1 \end{pmatrix}.
\end{equation*}
These equations and the relation \eqref{eq:A28} give
\begin{equation*}
T_{-} = \tilde{T}_{-} =  \frac{1}{\alpha_{-}(-L)} = \frac{1}{\overline{\alpha_+(L)}} \stackrel{\eqref{eq:A16}}{=}
T_{+},
\end{equation*} 
and 
\begin{equation*}
R_- = \frac{\gamma_-(-L)}{\alpha_-(-L)} = \frac{\overline{\gamma_+(L)}}{\overline{\alpha_+(L)}} 
\stackrel{\eqref{eq:A21}}{=}
\tilde R_{+},
\end{equation*} 
and 
\begin{equation*}
\tilde R_- = -\frac{\overline{\gamma_-(-L)}}{\alpha_-(-L)} = -\frac{{\gamma_+(L)}}{\overline{\alpha_+(L)}} 
\stackrel{\eqref{eq:A16}}{=}
R_{+},
\end{equation*} 
as stated in the lemma. $\Box$

\subsection{Proof of Theorem \ref{prop.1}}
\label{ap:1DAp.3}
Using the propagator matrices of the two random regions and the barrier, described in Appendices \ref{ap:1DAp.1}-\ref{ap:1DAp.2}, we have 
\begin{equation}
\begin{pmatrix}
\hat a(L) \\
\hat b(L) 
\end{pmatrix} = {\bf P}_+(L) {\bf P}_1{\bf P}_- (-L)\begin{pmatrix}
\hat a(-L) \\
\hat b(-L) 
\end{pmatrix},
\label{eq:3sect.1}
\end{equation}
To calculate the scattering matrix, we need a basic lemma. 

\vspace{0.05in}
\begin{lem}
\label{lema.A1}
Consider a system consisting of two successive sectors: The left one with propagator matrix ${\bf P}_l$ and the right one with propagator ${\bf P}_r$,
\begin{equation}
{\bf  P}_l = \begin{pmatrix} \alpha_l & \overline{\gamma_l} \\
\gamma_l & \overline{\alpha_l} 
\end{pmatrix} \quad 
\mbox{and} \quad {\bf  P}_r= \begin{pmatrix} \alpha_r & \overline{\gamma_r} \\
\gamma_r & \overline{\alpha_r} 
\end{pmatrix}.
\label{eq:A29}
\end{equation}
The propagator matrix of the system is 
$\displaystyle 
{\bf  P} = {\bf  P}_r {\bf  P}_l = \begin{pmatrix} \alpha & \overline{\gamma} \\
\gamma & \overline{\alpha}
\end{pmatrix}, $
where 
\begin{equation} 
\alpha = \alpha_l \alpha_r + \gamma_l \overline{\gamma_r}, \quad \gamma = \alpha_l \gamma_r + \gamma_l\overline{\alpha_r} .
\label{eq:A31}\end{equation}
The scattering matrix is $\displaystyle 
{\bf S} = \begin{pmatrix} T & \tilde R\\
R & T \end{pmatrix},
$
with entries
\begin{align}
T &= \frac{1}{ \overline{\alpha}} = 
T_{\rm l} T_{\rm r} (1-R_{\rm r} \tilde{R}_{\rm l})^{-1} , \label{eq:A33}\\
R &= - \frac{\gamma}{ \overline{\alpha}} =  R_{\rm l} + T_{\rm l}^2 R_{\rm r}  (1-R_{\rm r} \tilde{R}_{\rm l})^{-1} ,\label{eq:A34}\\
\tilde{R} &=  \frac{\overline{\gamma} }{ \overline{\alpha}} =  \tilde{R}_{\rm r} + T_{\rm r}^2 \tilde{R}_{\rm l}  (1-R_{\rm r} \tilde{R}_{\rm l})^{-1} .\label{eq:A35}
\end{align}
Here $T_j$, $R_j$ and $\tilde R_j$ are the transmission and reflection coefficients of the two sectors, with $j \in \{l,r\}$.

\end{lem}

\vspace{0.05in} \textbf{Proof:} Equation \eqref{eq:A31} follows trivially from the multiplication of the matrices 
\eqref{eq:A29}. The expression of the transmission and reflection coefficients in terms of $\alpha$ and $\beta$ 
is as in equations \eqref{eq:A16} and \eqref{eq:A21}. From definitions 
\begin{equation}
T_j = \frac{1}{\overline{\alpha_j}}, \quad R_j = - \frac{\gamma_j}{\overline{\alpha_j}}, \quad \tilde R_j = \frac{\overline{\gamma_j}}{\overline{\alpha_j}}, \quad j \in \{l,r\},
\label{eq:A36}
\end{equation}
we get that the transmission coefficient satisfies 
\begin{equation*}
T = \frac{1}{\overline{\alpha}} \stackrel{\eqref{eq:A31}}{=} \frac{1}{\overline{\alpha_l \alpha_r}} \left( 1 + \frac{\overline{\gamma_l}}{\overline{\alpha_l}} \frac{{\gamma_r}}{\overline{\alpha_r}}\right)^{-1}  \stackrel{\eqref{eq:A36}}{=} 
T_l T_r \left(1 - R_l \tilde R_r \right)^{-1}. 
\end{equation*}
For the reflection coefficient we have 
\begin{align*}
R &= -\frac{\beta}{\overline{\alpha}} \stackrel{\eqref{eq:A31}}{=} - \frac{\left(\alpha_l \gamma_r + \gamma_l \overline{\alpha_r} \right)}{
\overline{\alpha_l \alpha_r}} \left( 1 + \frac{\overline{\gamma_l}}{\overline{\alpha_l}} \frac{{\gamma_r}}{\overline{\alpha_r}}\right)^{-1} \\
&\stackrel{\eqref{eq:A36}}{=} \left(\frac{|\alpha_l|^2}{\overline{\alpha_l}^2} R_r + R_l \right]  \left(1 - \tilde R_l  R_r \right)^{-1} \\
&\stackrel{\eqref{eq:A17}}{=} \left[\frac{(1 + |\gamma_l|^2)}{\overline{\alpha_l}^2} R_r + R_l \right]  \left(1 - \tilde R_l R_r \right)^{-1} \\
&\stackrel{\eqref{eq:A36}}{=}  \left(T_l^2 R_r - R_l \tilde R_l R_r +  R_l \right)  \left(1 - \tilde R_l  R_r \right)^{-1} \\
&= T_l^2 R_r  \left(1 - \tilde R_l  R_r \right)^{-1} + R_l. 
\end{align*}
The derivation of the expression of the adjoint reflection coefficient is similar
\begin{align*}
\tilde R &= \frac{\overline{\beta}}{\overline{\alpha}} \stackrel{\eqref{eq:A31}}{=}  \frac{\left(\overline{\alpha_l \gamma_r} +\overline{ \gamma_l}\alpha_r \right)}{
\overline{\alpha_l \alpha_r}} \left( 1 + \frac{\overline{\gamma_l}}{\overline{\alpha_l}} \frac{\gamma_r}{\overline{\alpha_r}} 
\right)^{-1}  \\
&\stackrel{\eqref{eq:A36}}{=}  \left[ \tilde R_r + \frac{(1+|\gamma_r|^2)}{\overline{\alpha_r}^2} \tilde R_l 
\right]\left(1 - \tilde R_l  R_r \right)^{-1} \\
&\stackrel{\eqref{eq:A36}}{=}  \left( \tilde R_r + T_r^2 \tilde R_l - \tilde R_r R_r \tilde R_l \right)\left(1 - \tilde R_l  R_r \right)^{-1} \\
&= \tilde R_r + T_r^2 \tilde R_l \left(1 - \tilde R_l  R_r \right)^{-1}.   
\end{align*}
The proof of the lemma is complete. ~~$\Box$

To derive the expression of the transmission coefficient stated in Theorem \ref{prop.1}, we apply Lemma \ref{lema.A1} 
twice. The first time, we use the propagators ${\bf P}_l = {\bf P}_{-}(-L)$ and ${\bf P}_r = {\bf P}_1$ and obtain the transmission 
and reflection coefficients 
\begin{align}
T_{-,1} &\stackrel{\eqref{eq:A33}}{=} 
T_- T_{1} (1-R_{1} \tilde{R}_{-})^{-1} , \label{eq:T2sect}\\
R _{-,1} &\stackrel{\eqref{eq:A34}}{=}R_{-} + T_{-}^2 R_1  (1-R_1 \tilde{R}_{-})^{-1} ,\\\
\tilde{R}_{-,1} &\stackrel{\eqref{eq:A35}}{=} {R}_1 + T_1^2 \tilde{R}_{-}  (1-R_1 \tilde{R}_{-})^{-1} ,
\end{align}
with $T_-=T_-(-L)$, ${R}_{-}={R}_{-}(-L)$, and $\tilde{R}_{-}=\tilde{R}_{-}(-L)$.
Here we used that $R_1 = \tilde R_1$, according to equation \eqref{eq:transmB}. 
The second time we apply  Lemma \ref{lema.A1}, we use the propagators ${\bf P}_l = {\bf P}_{-,b}$ and ${\bf P}_r = {\bf P}_+(L)$. The transmission coefficient is 
\begin{align}
\mathcal{T} &\stackrel{\eqref{eq:A33}}{=} 
T_{-,1}T_{+}(1-R_{+} \tilde{R}_{-,1})^{-1}\nonumber \\ &\stackrel{\eqref{eq:T2sect}}{=} T_- T_1T_{+}
\left[ 1-R_1 \tilde{R}_{-}- R_+R_1(1-R_1 \tilde{R}_{-})- 
R_+ T_1^2 \tilde R_-
\right]^{-1}, \label{eq:Agen} 
\end{align}
with $T_+=T_+(L)$, ${R}_{+}={R}_{+}(L)$, and $\tilde{R}_{+}=\tilde{R}_{+}(L)$.
Now use the relations \eqref{eq:Tra2} in this equation to obtain 
\begin{align}
\label{eq:mathcalT}
\mathcal{T} &= T^2T_1\left[ 1 - 2 R R_1 + (R_1^2 - T_1^2) R^2 \right]^{-1}
\end{align}
and deduce from equation \eqref{eq:Reg2} that 
\begin{align}
R_1^2 - T_1^2= \frac{q^2 + 1}{(i + q)^2} = 2 R_1 - 1.
\label{eq:identRb}
\end{align}
The result \eqref{eq:Tra1} follows \eqref{eq:mathcalT} and  the identity
\begin{align*}
(1-R)\left[ 1 - \left(2R_1- 1\right) R\right] = 1 - 2 R R_1+ 
(2 R_1 - 1) R^2.
\end{align*}

We are interested in the mean transmitted intensity. To derive its expression, we recall from 
\cite[Section 7.1.1]{fouque2007wave} that 
$|R| < 1$. Since $R_1$ satisfies  equation \eqref{eq:Trab} and $R_1$ satisfies equation \eqref{eq:Trab}, 
we can use the series expansions
\begin{align*}
 (1 - R)^{-1} &= \sum_{k=0}^\infty R^k \quad \mbox{and} \quad
\big[1 - (2R_1-1 ) R \big]^{-1} = \sum_{k=0}^\infty 
(2R_1-1 )^k R^k,
\end{align*}
and rewrite equation \eqref{eq:Tra1} as
\begin{align}
\EE\big[ \big|\mathcal{T}\big|^2 \big] = |T_1|^2 \sum_{k_1,k_2,k_3,k_4 = 0}^\infty (2R_1-1 )^{k_2}
\big(2\overline{R_1}-1 \big)^{k_4}  \EE \left[ |T|^4 R^{k_1+k_2} \overline{R}^{k_3+k_4} \right]. \label{eq:Tra3}
\end{align}
It is shown in \cite[Chapters 7 and 9]{fouque2007wave} that 
\[
\EE \left[ |T|^2 R^{j} \overline{R}^{j'} \right] = 0, \quad \mbox{if}~ j \ne j',
\]
so only the terms with $k_1 + k_2 = k_3 + k_4$ 
contribute in \eqref{eq:Tra3}. Moreover, since 
$
|R|^2 = 1 - |T|^2$, we obtain that 
\begin{align}
\EE\big[ \big|\mathcal{T}\big|^2 \big] = |T_1|^2 \sum_{k=0}^\infty \sum_{k_2,k_4 = 0}^k (2R_1-1 )^{k_2}
\big(2\overline{R_1}-1 \big)^{k_4}  \EE \left[ |T|^4\big(1 - |T|^2 \big)^k \right]. \label{eq:Tra4}
\end{align}
Now, use the notation \eqref{eq:Tra5}
and observe that 
\begin{align*}
|T_1|^2 \sum_{k_2,k_4 = 0}^k (2R_1-1 )^{k_2}
\big(2\overline{R_1}-1 \big)^{k_4}  &= \left| T_1 \sum_{k_2=0}^k (2R_1-1)^{k_2} \right|^2 \\
&= \left| \frac{T_1}{2( 1- R_1)} \left[ 1-(2R_1-1)^{k+1} \right]\right|^2,
\end{align*}
where according to equation \eqref{eq:Reg2} we have 
\[
\left| \frac{T_1}{2(1- R_1)} \right|^2 = \frac{1}{4}.
\]
The result \eqref{eq:meanint} follows, once we recall the definition \eqref{eq:Tra5} of $\tau_k$. 
$ \Box$

\subsection{Transmission through two independent random sections}
\label{ap:1DAp.4}
To derive the mean transmitted intensity in the absence of symmetry, we begin with the general formula 
\eqref{eq:Agen}, where now the transmission and reflection coefficients in the two random sections are statistically 
independent. Using equation \eqref{eq:identRb} in \eqref{eq:Agen} and writing the inverse of the curly bracket 
as power series, we get 
\begin{align*}
\EE\big[ |{\cal T}|^2\big] = |T_1|^2 
\sum_{j,l=0}^\infty 
\EE\Big[ |T_-|^2|T_+|^2 
\big[ R_1 (R_++\tilde{R}_-) +(1-2R_1) R_+ \tilde{R}_- \big]^j
\\
\times \overline{\big[R_1({R_+}+{\tilde{R}_-}) +(1-2{R_1}) {R_+} {\tilde{R}_-} \big]}^l
\Big].
\end{align*}
Next, we expand the $j$ and $l$ powers using the binomial theorem and use the independence of $(T_-,\tilde R_-)$ and 
$(T_+,\tilde R_+)$. Using also that  $\EE[|T_+|^2 R_+^n\overline{R_+}^m]=0$ unless $m=n$, and the 
same for $(T_-, \tilde{R}_-)$, we get the result \eqref{eq:meanIndep}. 

\section{Derivation of the results for random waveguides}
\label{ap:WavegAp}
In this appendix we prove the results stated in section \ref{sect:Waveg}. The frequency $\om$ is fixed, so we simplify notation as in the previous appendix, by droping the $\om$ argument. 
\subsection{Proof of Theorem \ref{thm.W1}}
\label{ap:WavegAp.1}
We obtain from equations (\ref{eq:Wa40}-\ref{eq:Wa41}) that 
\begin{equation}
{\bf P}_+^{(a)} = {\bf T}_+ \left( \bI - \overline{{\bf R}_+} {\bf R}_+\right)^{-1}  \quad \mbox{and} \quad 
{\bf P}_+^{(b)} = -\overline{{\bf T}_+} \left( \bI -{\bf R}_+\overline{ {\bf R}_+}\right)^{-1} {\bf R}_+.
\label{eq:B02}
\end{equation}
Moreover, standard formulas for block matrix inversion give  that 
\begin{equation}
{{\bf P}^{-1}_-(-L)} \stackrel{\eqref{eq:Wa39}}{=}
\overline{{\bf P}^{-1}_+(L)} = \begin{pmatrix} 
\overline{{\bf T}^{-1}_+} & {\bf R}_+ {\bf T}_+^{-1} \\ \\
\overline{{\bf R}_+ {\bf T}_+^{-1}} & {\bf T}_+^{-1} 
\end{pmatrix}.
\label{eq:B03}
\end{equation}
Then, using this result in  \eqref{eq:Wa47} and recalling the block algebraic structure of 
${\bf P}_+$ and ${\bf P}_1$, we get that 
the propagator of the system has the form
\begin{equation}
\cbP = \begin{pmatrix} 
\cbP^{(a)} & \overline{\cbP^{(b)}} \\
\cbP^{(b)} & \overline{\cbP^{(a)}}
\end{pmatrix} .
\label{eq:B04}
\end{equation}
We are interested in the first block $\cbP^{(a)}$, which according to definition \eqref{eq:Wa48} defines the transmission matrix 
\begin{equation}
\cbT = \left[\, \overline{\cbP^{(a)}}\, \right]^{-1}.
\label{eq:B05}
\end{equation}
The expression of this block follows by carrying out the multiplication in \eqref{eq:Wa47},
\begin{align*}
\cbP^{(a)} = {\bf T}_+ \left( \bI - \overline{{\bf R}_+} {\bf R}_+\right)^{-1} 
\left({{\bf P}_1^{(a)}} - 
\overline{{\bf R}_+}{\bf P}_1^{(b)} + \overline{{\bf P}_1^{(b)}{\bf R}_+}  - 
\overline{{\bf R}_+{\bf P}_1^{(a)} {\bf R}_+} \right) \left(\overline{{\bf T}_+}\right)^{-1}.
\end{align*}
But we also have from the relations (\ref{eq:Wa42}-\ref{eq:Wa43}) that 
\begin{align*}
\bI - \overline{{\bf R}_+} {\bf R}_+ \approx \bI -{\bf R}^{\star}_+ {\bf R}_+ 
= {\bf T}^{\star}_+ {\bf T}_+,
\end{align*}
which simplifies the factor
\begin{align}
{\bf T}_+ \left( \bI - \overline{{\bf R}_+} {\bf R}_+\right)^{-1} \approx  {\bf T}_+ \left({\bf T}^{\star}_+ {\bf T}_+ \right)^{-1} =  \left({\bf T}^{\star}_+ \right)^{-1}. \label{eq:B1Approx}
\end{align}
The statement of the theorem follows from \eqref{eq:B05} and the relations 
\[
{\bf P}_1^{(a)} = \overline{{\bf T}_1^{-1}}, \quad {\bf P}_1^{(b)} = - \overline{{\bf P}_1^{(b)}} = - 
{\bf T}_1^{-1} {\bf R}_1,
\]
deduced from equations \eqref{eq:Wa15} and (\ref{eq:Wa26}-\ref{eq:Wa27}). $~~ \Box$

\subsection{Proof of Theorem \ref{thm.W2}}
\label{ap:WavegAp.2}
Weak scattering in the random medium means that the norm of the reflection matrix ${\bf R}_+$ is small.
Thus, we can use Neumann series to approximate the square bracket in \eqref{eq:Wa49} by 
\begin{align}
{\bf Q} &= \left[ {\bf T}_1^{-1} - {\bf R}_+ {\bf T}_1^{-1} { \bf R}_1 - {\bf T}_1^{-1}{ \bf R}_1{\bf R}_+ - 
{\bf R}_+ \overline{{\bf T}_1^{-1}} {\bf R}_+ \right]^{-1} \nonumber \\ 
&= \left[ \bI - {\bf T}_1 {\bf R}_+ { \bf R}_1 {\bf T}_1^{-1} - { \bf R}_1{\bf R}_+ - 
{\bf T}_1{\bf R}_+ \overline{{\bf T}_1^{-1}} {\bf R}_+ \right]^{-1}{\bf T}_1 \nonumber \\
&\approx {\bf T}_1 + {\bf T}_1 {\bf R}_+ {\bf R}_1 + {\bf R}_1 {\bf R}_+ {\bf T}_1,
\label{eq:B5}
\end{align}
where in the second equality we used that ${\bf R}_1$ and ${\bf T}_1^{-1}$ commute, because they are diagonal.
{This approximation is valid for weak scattering and neglects terms that contain a product involving two (or more) reflection matrices ${\bf R}_+$.}
Substituting \eqref{eq:B5} into \eqref{eq:Wa49}, we get that 
\begin{align}
\cbT \approx {\bf T}_+ \left( {\bf T}_1 + {\bf T}_1 {\bf R}_+ {\bf R}_1+ {\bf R}_1 {\bf R}_+ {\bf T}_1 \right) 
{\bf T}_{+}^{T},
\label{eq:B6}
\end{align}
and the mean transmittivity is, from \eqref{eq:Wa50},
\begin{align}
\EE \left[ \sum_{j,l = 1}^N | \cT_{jl}|^2 \right] \approx \mbox{Tr} \left\{ \EE \left[  
\overline{\left(\bI - {\bf R}_+^{\star} {\bf R}_+ \right)} \left({\bf T}_1 + {\bf T}_1 {\bf R}_+ {\bf R}_1+ {\bf R}_1 {\bf R}_+ {\bf T}_1 \right)^\star
 \right. \right. \nonumber \\
\left. \left. \times \left(\bI - {\bf R}_+^{\star} {\bf R}_+ \right)\left({\bf T}_1 + {\bf T}_1 {\bf R}_+ {\bf R}_1+ {\bf R}_1 {\bf R}_+ {\bf T}_1 \right)
\right] \right\}.
\label{eq:B7}
\end{align}
Here we used the energy conservation relation \eqref{eq:Wa42} and the commutation property  of the trace
\begin{equation*}
\mbox{Tr}\left[  
\overline{{\bf T}_+} {\bf A} {\bf T}_+^{T} \right] = 
\mbox{Tr}\left[  
{\bf T}_+^{T} \overline{{\bf T}_+} {\bf A}  \right] 
\stackrel{\eqref{eq:Wa42}}{=} \mbox{Tr}\left[  
\left(\bI - {\bf R}_+^{T} \overline{{\bf R}_+}\right) {\bf A}  \right], \qquad \forall \, {\bf A} \in \mathbb{C}^{N \times N}.
\end{equation*}
{The approximation (\ref{eq:B7}) is consistent with  (\ref{eq:B5}) because, if $n \ne n'$, $n , n' \ge 0$, then
\[
 \EE \left[ \prod_{k=1}^n {R}_{+,j_k l_k} \prod_{k'=1}^{n'}   \overline{{R}_{+,j'_{k'}l'_{k'}}} \right] = 0 ,  
\]
for any $j_k,l_k,j'_{k'},l'_{k'} \in \{1,\ldots,N\}$,
as shown by the analysis of the statistical moments of the transmission and reflection matrices of the random medium given in 
\cite{garnier2008effective}. This is why we could neglect the quadratic terms in (\ref{eq:B5}). Only the terms that do not involve ${\bf R}_+$ or that involve two reflection matrices, with one of them being complex-conjugated,
contribute to the approximation of \eqref{eq:B7}.
Thus, the mean transmissivity is  approximated by} 
\begin{align}
\TT = &  \mbox{Tr} \left\{ \EE \left[  \overline{\left(\bI - {\bf R}_+^{\star} {\bf R}_+ \right)} {\bf T}_1^\star {\bf T}_1 - 
 {\bf T}_1^\star {\bf R}_+^{\star} {\bf R}_+ {\bf T}_1+ {\bf R}_1^\star {\bf R}_+^{\star} {\bf T}_1^\star {\bf T}_1
 {\bf R}_+ {\bf R}_1
\right. \right.  \nonumber \\
&\left. \left. + {\bf R}_1^\star {\bf R}_+^{\star} {\bf T}_1^\star {\bf R}_1 {\bf R}_+ {\bf T}_1 + 
{\bf T}_1^\star {\bf R}_+^{\star}{\bf R}_1^\star {\bf T}_1 {\bf R}_+ {\bf R}_1 + {\bf T}_1^\star 
{\bf R}_+^{\star}{\bf R}_1^\star {\bf R}_1  {\bf R}_+ {\bf T}_1
\right] \right\}.
\end{align}
The statement of the theorem follows from this equation once we write explicitly the trace and use the expressions 
(\ref{eq:Wa26}-\ref{eq:Wa27}) of the entries of ${\bf T}_1$ and ${\bf R}_1$. 

\subsection{Proof of Proposition \ref{prop.W1}}
\label{ap:WavegAp.3}
The propagator matrix of the waveguide system with two independent random sections is 
\begin{equation}
\cbP = {\bf P}_+(L) {\bf P}_1 \check{{\bf P}}_+(L),
\label{eq:B10}
\end{equation} 
where $\check{{\bf P}}_+$ is an independent and identically distributed copy of ${\bf P}_+$.
Given the algebraic structure 
of the propagator ${\bf P}_1$ of the barrier given in \eqref{eq:Wa14}, and of the random medium propagator 
${\bf P}_+$ given in equations \eqref{eq:Wa31} and \eqref{eq:Wa39}, we conclude from \eqref{eq:B10} that 
$\cbP$ is of the form \eqref{eq:B04}. We are interested in its first block $\cbP^{(a)}$ which determines the transmission 
matrix $\cbT$, as in equation \eqref{eq:B05}. 

Using equation \eqref{eq:B02} and multiplying through in equation \eqref{eq:B10} we get that 
\begin{align*}
\cbP^{(a)} = {\bf T}_+ \left( \bI - \overline{{\bf R}_+} {\bf R}_+ \right)^{-1} 
\left[ \left({\bf P}^{(a)}_1 - \overline{{\bf R}_+} {\bf P}^{(b)}_1\right) \check{\bf T}_+\left( \bI - \overline{\check{\bf R}_+} \check{\bf R}_+ \right)^{-1} \right. \\
\left. -  \overline{\left({\bf P}^{(b)}_1 - {\bf R}_+ {\bf P}^{(a)}_1\right)}
\overline{\check{\bf T}_+}\left( \bI - {\check{\bf R}_+} \overline{\check{\bf R}_+} \right)^{-1} \check{\bf R}_+
\right],
\end{align*}
where the first factor is approximated in \eqref{eq:B1Approx}. 
This gives 
\begin{align} 
\cbT = \left(\overline{\cbP^{(a)}} \right)^{-1} \approx &\left[ \left(\overline{{\bf P}^{(a)}_1} - {{\bf R}_+} \overline{{\bf P}^{(b)}_1}\right) \overline{\check{\bf T}_+}\left( \bI - {\check{\bf R}_+} \overline{\check{\bf R}_+} \right)^{-1} \right. \nonumber\\
&\left. -  {\left({\bf P}^{(b)}_1 - {\bf R}_+ {\bf P}^{(a)}_1\right)}
{\check{\bf T}_+}\left( \bI - \overline{\check{\bf R}_+}{\check{\bf R}_+} \right)^{-1} \overline{\check{\bf R}_+}
\right]^{-1} {\bf T}_+^{T},
\end{align}
where the square bracket can be approximated with Neumann series for small reflection matrices. Such series are also 
used to expand 
$\mbox{Tr} (\cbT^{\star} \cbT )$  up to second order in terms of the reflection matrices of the random medium. The result stated in the proposition follows after we take the expectation and use that 
$( {\bf T}_+, {\bf R}_+)$ and $( \check{\bf T}_+, \check{\bf R}_+)$ are independent.

\bibliographystyle{siam} 
\bibliography{biblio.bib}

\end{document}